\documentclass[aps,prb,twocolumn,superscriptaddress,dvipsnames,longbibliography, floatfix]{revtex4-2}

\usepackage[utf8]{inputenc}
\usepackage{amssymb,amsfonts,amsmath} 
\usepackage{graphicx,epsfig,psfrag}
\usepackage{color}
\usepackage{bbm}
\usepackage{natbib}
\usepackage{mathtools}
\usepackage{esint}
\usepackage[caption=false]{subfig}
\usepackage{epstopdf}
\usepackage{comment}
\usepackage{upgreek}
\usepackage{hyperref}
\hypersetup{
    colorlinks = true,
    linkcolor  = blue,
    citecolor  = blue,
    urlcolor   = blue,
    linktocpage= true
}
\begin{document}

\title{Anderson localization of emergent quasiparticles: Spinon and vison interplay at finite temperature in a \texorpdfstring{$\mathbb{Z}_2$}{} gauge theory in three dimensions}

\author{Minho Kim}
\affiliation{TCM group, Cavendish Laboratory, University of Cambridge. Cambridge CB3 0HE, United Kingdom}
\author{Giuseppe De Tomasi}
\affiliation{Department of Physics, University of Illinois at Urbana-Champaign, Urbana, Illinois 61801-3080, USA}
\author{Claudio Castelnovo}
\affiliation{TCM group, Cavendish Laboratory, University of Cambridge. Cambridge CB3 0HE, United Kingdom}

\begin{abstract}
Fractional statistics of quasiparticle excitations often plays an important role in the detection and characterization of topological systems. In this paper, we investigate the case of a three-dimensional (3D) $\mathbb{Z}_2$ gauge theory, where the excitations take the form of bosonic spinon quasiparticle and vison flux tubes, with mutual semionic statistics. We focus on an experimentally relevant intermediate temperature regime, where sparse spinons hop coherently on a dense quasistatic and stochastic vison background. The effective Hamiltonian reduces to a random-sign bimodal tight-binding model, where both the particles and the disorder are borne out of the same underlying quantum spin liquid (QSL) degrees of freedom, and the coupling between the two is purely driven by the mutual fractional statistics. We study the localization properties and observe a mobility edge located close to the band edge, whose transition belongs to the 3D Anderson model universality class. Spinons allowed to propagate through the quasistatic vison background appear to display quantum diffusive behavior. When the visons are allowed to relax, in response to the presence of spinons in equilibrium, we observe the formation of vison depletion regions slave to the support of the spinon wavefunction. We discuss how this behavior can give rise to measurable effects in the relaxation, response and transport properties of the system and how these may be used as signatures of the mutual semionic statistics and as precursors of the QSL phase arising in the system at lower temperatures.  \end{abstract}

\maketitle
%
%

\section{\label{sec:intro}
Introduction
        } 
Quantum spin liquids~\cite{Savary_2016,Zhou_2017, Knolle_18, Broholm_2020} (QSLs) are intriguing phases of matter characterized by highly correlated spins that fail to order in a conventional way down to temperatures much lower than the characteristic interaction energy scale in the system. This leads to the appearance of exotic properties, including emergent gauge symmetries, fractionalized excitations with anyonic statistics, and topological order~\cite{Xiao_2007, Wilczek_09, moessner_moore_2021}.

Such phenomenology is both of fundamental interest and has potential applications in quantum information storage and processing~\cite{Kitaev2003, Stringnet, Nayak_08}. Despite the growing number of candidate spin liquid materials discovered in laboratories in recent years, their experimental detection and characterization remains a paramount challenge to date~\cite{Broholm_2020}. 

The fractional behavior of the quasiparticle excitations in QSLs is often considered a promising handle to make them apparent in experiments~\cite{Han_12, Nasu_16, Yoshitake_16, Shen_16, Paddison_17, Morampudi_17, Yoshitake_17, Yoshitake_17_2, Do_17, Chatterjee_19, Halasz_19, MainProp, MainCorr}. Authors of recent work~\cite{MainProp, QSLConstruction, MainCorr} focused on a class of systems --- encompassing spin ice~\cite{Gingras_2014, Udagawa_21}, Kitaev materials~\cite{KitaevMaterials1}, and valence bond systems~\cite{Moessner_2011} --- where the QSL behavior is borne out of the interplay of some large energy scale that acts projectively on the Hilbert space --- inducing on its own a classical spin liquid state~\cite{Castelnovo_07} --- and smaller transverse (i.e., kinetic) terms~\cite{Balents_02, Hermele_04}. 
Considering two-dimensional (2D) models, the presence of a finite temperature regime was brought to light, where the anyonic statistics between the quasiparticles leads to a rich phenomenology driven by Anderson localization~\cite{Anderson_58} of emergent tight binding particles, where the disorder is itself emergent from the same spin degrees of freedom in an otherwise disorder-free system~\cite{Schiulaz_15, Yao_16, Smith_17, Yarloo_18, smith2019thesis}. The resulting effects on relaxation and transport properties~\cite{MainProp, MainCorr} are both of fundamental interest in their own right as well as of potential experimental interest as precursor diagnostics tools and bellwethers for QSL phases at lower temperatures.

In this paper, we revisit the intriguing finite-temperature juncture of topological behavior and Anderson localization of emergent quasiparticles in disorder-free systems by extending it to three dimensions (3D). This brings about a few important changes, whose study is the key interest of this paper. On the one hand, nontrivial mutual statistics in dimensions higher than 2D requires at least one of the quasiparticle species to take the form of an extended rather than pointlike object (e.g., a closed loop, in our system of choice). On the other hand, 3D is the smallest dimension in which the system has a metal-insulator Anderson transition as a function of energy, producing a so-called mobility edge~\cite{Evers_2008} that is absent in the 2D model considered earlier~\cite{MainProp,MainCorr}.

Specifically, we investigate a $\mathbb{Z}_2$ topological spin liquid system on the cubic lattice~\cite{3DToric, Toric3DCastelnovoChamon}, which exhibits two species of fractionalized excitations (spinons and visons) with mutual semionic statistics. The spinons are pointlike (i.e., end points of open strings), and we assume their energy cost to be the largest term in the Hamiltonian, with their transverse or hopping term as the first subleading energy scale. As a first approximation, the spinons can effectively be modeled as sparse tight-binding particles on the lattice. The visons take instead the form of closed loops (i.e., edges of open membranes embedded in 3D) and we assume their energy cost as well as transverse or hopping terms to be negligibly small (effectively vanishing) compared with the temperature. As such, the visons can effectively be modeled as classical stochastic objects that are quasistatic on the characteristic transport and equilibration timescales of the spinons. The visons therefore act as a stochastic background that affects the behavior of the spinons through mutual semionic statistics --- a behavior that is equivalent to charges moving on a lattice where plaquettes are randomly threaded by half quantum ($\pi$) fluxes~\cite{MainProp}. Vice versa, the spinons endow the vison configuration with an energy that can affect their correlations in an important way~\cite{MainCorr}. 

In the first instance, we study the localization properties of a spinon moving across a stochastic and static vison background. This is equivalent to an electron hopping on a cubic lattice with randomly distributed $\pi$ fluxes threading its plaquettes --- a model that has hitherto not been considered in the literature. 
Whereas in 2D all states are localized away from the middle of the band, we find clear evidence of a mobility edge in 3D, which occurs very close to the band edge. We find that the Anderson transition of our model belongs to the same universality class as the orthogonal 3D Anderson model~\cite{Andersonscale1, Andersonscale2}, and the states inside the mobility edges are indeed extended.

We then proceed to study the behavior of spinons and visons, following the path set by Refs.~\onlinecite{MainProp, MainCorr}. Despite the differences in the properties of the eigenstates in 3D, the behavior we observe is very similar to the 2D case. We find slightly weaker localization effects, as reflected by the fact that the spinon propagation behavior in the $\pi$ flux background in 3D agrees well with the diffusive like continuous random flux case --- unlike in 2D where the propagation showed a remarkable departure toward anomalous diffusion~\cite{MainProp}. Notwithstanding, once the visons are allowed to move and relax stochastically in the presence of the spinons, we do observe once again the formation of the characteristic depletion regions~\cite{MainCorr} that are responsible for the predicted anomalous relaxation and transport effects. This is once again in stark contrast with the case where mutual statistics is replaced by interactions between spinons and visons, where the formation of such depletion regions is not observed. 

Our results show that the rich phenomenology arising at the interface between topological order and Anderson localization in disorder-free systems is not limited to low-dimensional systems. Moreover, while the localization properties are different, the notable effects on the relaxation and transport properties underpinned by the formation of vison depletion regions around the spinons survive. This is once again a behavior due to the mutual semionic statistics and, if observed, can thus be taken as a signature of 3D QSL behavior at finite temperature. 

While we must wait for the discovery of candidate materials that realize $\mathbb{Z}_2$ QSL phases in 3D, our results may be relevant to other contexts, including frustrated magnetic pyrochlore oxides and resonant valence bond systems. Moreover, the possibility of realizing $\mathbb{Z}_2$ spin liquid Hamiltonians in our temperature regime with quantum annealers~\cite{Boothby_19, QSLConstruction, QuantumAnnealExp1} and quantum simulators~\cite{Simulator1,Simulator2} could provide a suitable arena where the physics discussed here could be tested and explored further. 

The outline of the paper is as follows. In Sec.~\ref{sec:model}, we introduce the 3D model and its Hamiltonian as well as the assumptions made for the temperature regime of interest in this paper. In Sec.~\ref{sec:mobility}, we discuss the localization properties and verify the existence of a mobility edge in the system. We then proceed to study the diffusion of spinons in a static random vison background in Sec.~\ref{sec:propagation}, and we compare it with analytical results based on a Bethe lattice approximation~\cite{MainCorr}. In Sec.~\ref{sec:spinonvison}, we allow the visons to relax stochastically while the spinons remain instantaneously in thermodynamic equilibrium. As in 2D, we observe the formation of mutual-semionic-statistics-induced vison depletion regions around each spinon. We describe the effects that this behavior has on the response and transport properties of the system, particularly on the out-of-equilibrium dynamics when the temperature is varied quickly enough, in Sec.~\ref{sec:out-of-equilibrium}. Finally, we present our conclusions and outlook in Sec.~\ref{sec:conclusions}. 
%
%

\section{\label{sec:model}
Model} 
\begin{figure}
    \centering
    \subfloat{
		\includegraphics[width=0.47\linewidth]{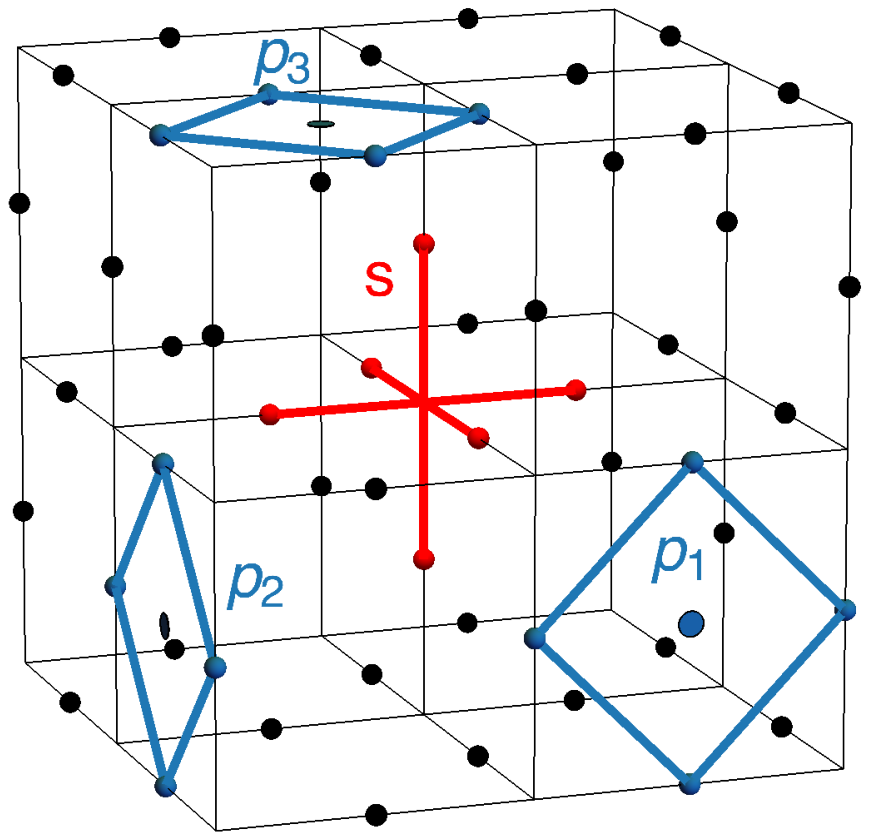}
	}
    \subfloat{
    \includegraphics[width=0.47\linewidth]{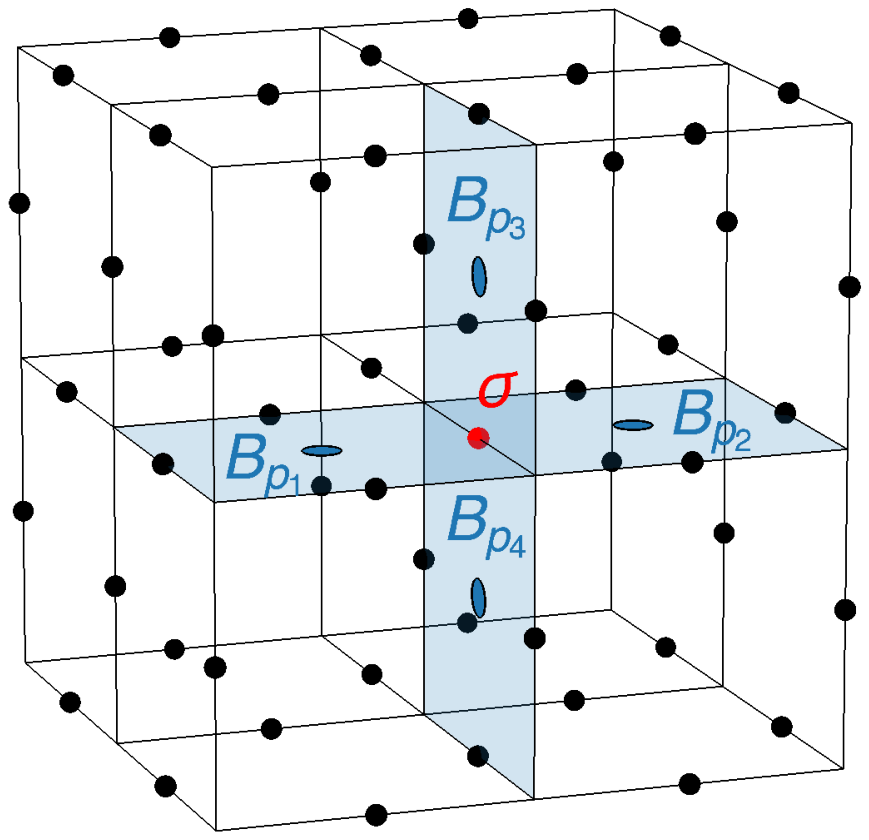}
    }
    \caption{\label{fig:modelvis}
    Visualization of the lattice system considered in this paper, with spin-$\frac{1}{2}$ degrees of freedom living on the bonds of a cubic lattice (black solid dots). 
    Star and plaquette operators at site $s$ and plaquettes $p_1, p_2, p_3$, respectively, are shown for illustration purposes (left panel).
    Acting on any given spin (e.g., the red dot labeled $\sigma$ in the right panel) can affect the state of the two adjacent star operators (not shown), and/or the state of the four adjacent plaquette operators ($B_{p_1}$, $B_{p_2}$, $B_{p_3}$, and $B_{p_4}$ in the figure).} 
\end{figure}

In this section we introduce the 3D toric code model~\cite{3DToric, Toric3DCastelnovoChamon} (see also Appendix D in Ref.~\onlinecite{CCS}) and the corresponding effective quasiparticle Hamiltonian. 
The model is defined on an $L^3$ cubic lattice with periodic boundary conditions. Spin-$\frac{1}{2}$ degrees of freedom $\sigma_i$ live on the bonds of the lattice, labeled by the index $i = 1, 2, \cdots, 3L^3$ (see Fig.~\ref{fig:modelvis}). 
We further label the elementary square plaquettes on the lattice with the index $p = 1, 2, \cdots, 3L^3$, and the lattice sites with $s = 1, 2, \cdots, L^3$. 

We introduce the plaquette and star operators:
\begin{equation}
    B_p = \prod_{i \in p} \sigma_i^z 
    \qquad {\rm and} \qquad 
    A_s = \prod_{i \in s} \sigma_i^x
    \, , 
\end{equation}
where for convenience of notation $i \in p$ and $i \in s$ indicate the four spins around a plaquette $p$ and the six spins adjacent to a site $s$, respectively. Examples of spin and plaquette operators are shown in Fig.~\ref{fig:modelvis}. 

The Hamiltonian reads
\begin{equation}
    H = -\frac{\Delta_s}{2} \sum_s A_s 
    -\frac{\Delta_v}{2} \sum_p B_p
    \, , 
\label{eq:TCham}
\end{equation}
where $\Delta_s, \Delta_v > 0$. 
All plaquette and star operators commute with one another, and the ground state of the toric code Hamiltonian is obtained when they all have eigenvalue $+1$. Negative eigenvalues of $A_s$ and $B_p$ label the excited states. Both types of excitations are gapped; the former are dubbed star defects or spinons (energy cost $\Delta_s$), whereas the latter are dubbed plaquette defects or visons (energy cost $\Delta_v$). 
Notice that both types of excitations are static according to Eq.~\eqref{eq:TCham}. In the following, it will be useful to introduce a vison occupation variable $n_p = 0,1$ [namely, $n_p = (1 - B_p)/2$]. 

Changing the sign of the $\sigma^x$ component of a spin on a given bond $i$ results in flipping the sign of the expectation values of the pair of star operators on the $s,s'$ sites adjacent to $i$. This amounts to the creation/annihilation of a pair of spinons or to the hopping of a spinon between the two sites $s$ and $s'$, depending on the local initial state of the spins. This directly relates to the well-known gauge symmetry in the $\sigma^x$ degrees of freedom, whereby spinons act as sinks/sources of the gauge field. 

Similarly, changing the sign of the $\sigma^z$ component of a spin on a given bond $i$ results in flipping the sign of the expectation values of the four plaquette operators on the plaquettes adjacent to $i$ (see Fig.~\ref{fig:modelvis}, right panel). This leads to an equivalent gauge structure that is, however, formulated in a redundant way, with an extensive number of conserved quantities~\cite{Toric3DCastelnovoChamon}.
As a consequence of the cooperative flipping of the eigenvalues of plaquette operators, visons must necessarily form edges of open 2D membranes embedded in the dual 3D cubic lattice (see Refs.~\onlinecite{3DToric, Toric3DCastelnovoChamon} for details). This is indeed a key feature that allows spinons and visons to exhibit nontrivial mutual statistics in 3D, as the operation of winding a spinon around a vison remains well defined: as a spinon moves along a closed loop on the cubic lattice, it either does or does not cross a given vison membrane (an odd number of times); if it does, the final state acquires an overall minus sign with respect to the starting one --- spinons and visons obey mutual semionic statistics. On their own, instead, both spinons and visons are bosonic excitations. 

In this paper, we are interested in a modification of the toric code Hamiltonian in Eq.~\eqref{eq:TCham}, where a longitudinal field $h \sum_i \sigma^z_i$, $h < \Delta_s$, is added and the vison gap $\Delta_v$ is taken to be small enough to be negligible (e.g., much smaller than some thermal energy scale $0 < T \ll \Delta_s$). The effects of the field are two fold: (i) spinon excitations become itinerant and form a dispersive band, and (ii) the vison gap gets perturbatively renormalized (which is a small effect and does not conflict with the assumption above that it is negligible, at least at small but finite temperatures). For a justification of these assumptions in relation to experimentally relevant models of spin liquids, we refer the reader to the introductory sections in Refs.~\onlinecite{MainProp,MainCorr}. 

In this regime, the spinons behave as quantum itinerant quasiparticles that can be effectively described in the sparse limit by a tight-binding Hamiltonian~\cite{MainProp}. The visons, on the other hand, remain static. 
Given the mutual semionic statistics between spinons and visons~\cite{Kitaev2003}, the presence of the latter means that the spinon wavefunction in the effective tight-binding model picks up a minus sign as the spinon hops around a closed loop on the lattice traversing a vison membrane an odd number of times. This is equivalent to the behavior of a unit charge moving around a loop threaded by half an elementary magnetic flux unit ($\pi$ flux). 
The vison occupancy $n_p = 0,1$ defines the $\pi$ flux pattern through which spinons move. 
It is therefore convenient to rewrite the effective Hamiltonian as
\begin{equation}
    \label{eqn:MainHamiltonian}
    H_{\textrm{eff}} = 
    \Delta_s \sum_s b_s^{\dagger} b_s 
    - \eta_s \sum_{\langle s, s' \rangle} \left[ \exp (i A_{s s'}) b_s^\dagger b_{s'}
    + {\rm H.c.}\right]
\, , 
\end{equation}
where $(\nabla \times A)_p = \Phi_p = \pi \, n_p$ and $A_{ss'} = -A_{s's}$. Here, $b_s, \; b_s^\dagger$ are the spinon annihilation and creation operators at lattice site $s$, which obey bosonic statistics. The second sum runs over pairs of nearest-neighbor lattice sites, and we relabeled the field $h \to \eta_s$ for notational convenience in the new language. 
One should further constrain the spinon density operator $b_s^{\dagger} b_s$ to take on values $0,1$ only and include spinon annihilation and creation terms. However, in this paper we will primarily concern ourselves with the behavior of isolated spinons in the low spinon density limit. Therefore, we will safely ignore the constraint as well as the additional terms. 

Generally speaking, one may expect small perturbations in the system to give rise to hopping amplitudes for visons as well (the visons ought to move subject to the gauge constraint discussed above, which is not equally amenable to a pointlike particle hopping description, as is the case for the spinons). In this paper we assume that such terms are small compared with the already small vison gap $\Delta_v$ (otherwise, they would drive the system out of the topological phase that underpins the physics of interest in this paper) and hence also small compared with temperature. In this regime, coherent vison dynamics is unlikely to survive on timescales of significance to the results discussed in this paper; therefore, where appropriate, we shall model the vison dynamics stochastically in a Markovian approximation (namely, using Monte Carlo (MC) simulations that respect the gauge structure of the system). Moreover, we are justified to take a Born-Oppenheimer perspective and approximate the visons to be quasistatic on the timescale of motion/equilibration of the spinons; namely, at short times, we can assume the visons to be static (Sec.~\ref{sec:propagation}) with respect to spinon motion, whereas at long times, we can assume the spinons to be instantaneously in equilibrium on the timescales over which the vison stochastic dynamics occurs (Sec.~\ref{sec:spinonvison}).
%
%

\section{\label{sec:mobility}
Energy eigenstate localization properties}
As discussed in Sec.~\ref{sec:model}, see, e.g., Eq.~\eqref{eqn:MainHamiltonian}, visons introduce random $\pm$ phases in the spinon hopping amplitudes $\eta_s$. One can therefore draw a parallel with random-hopping tight-binding particles in 3D lattices. However, exploration of the case of bimodal hopping ($\pm \eta_s$) where the distribution of signs is induced by an emergent $\mathbb{Z}_2$ gauge structure is lacking. For this reason, we include here a brief study of the localization properties of the energy eigenstates of our system while leaving a more thorough characterization thereof to future work. 

Anderson localization with off-diagonal disorder is somehow unique~\cite{Dyson_53,Weissmann_1975,Fleishman_1977, Ziman_82, Soukoulis_82,CainRomerSchreiber,Brouwer_2000, Biswas2000, Xiong_2001, Evangelou2003, Taraskin2002, Nikolic2001, Eilmes2001}. The presence of chiral (sublattice) symmetry leads to a divergence of the density of states (DOS) at the band center ($E\approx 0$). Consequently, the eigenstates around the divergence might behave differently from all others. For instance, in one dimension, all eigenstates away from $E\approx 0$ are exponentially localized~\footnote{The case of random phases is special in one dimension. Indeed, using gauge transformation $b_s-> \exp(i\phi_s) b_s$ one can absorb the phase.}. However, around $E\approx 0$, there exists a divergence in the DOS which scales as $\rho(E)\sim \frac{1}{E\ln^3{E}}$~\cite{Dyson_53,Brouwer_2000}, and the eigenstates with $E\approx 0$ are only weakly localized~\cite{Fleishman_1977, Soukoulis_81,Ziman_82} with $\psi(E\approx 0) \sim \exp(-\gamma\sqrt{r})$~\footnote{Formally, the localization length is infinite.}. In higher dimensions, the question of localization at the band center is still unresolved. At large enough disorder, a general argument predicts $\psi(E \approx 0) \sim \exp({-\gamma \sqrt{\ln{r}}})$~\cite{Ziman_82}. However, finite-sized numerical simulations found the existence of power law localized states~\cite{Xiong_2001}.

In addition to the localization behavior for states around $E\approx 0$, the general belief is that off-diagonal systems are like the ones with onsite disorder. Thus, the critical dimension is still $D=2$, and for larger dimensions, one expects an Anderson transition as a function of disorder strength and the existence of mobility edges within the extended phase. In our case, the fluctuations of the magnitude of the disorder is fixed, i.e., $|\exp{i A_{s,s'}}|=1$. This has two immediate consequences: first, we do not have a tuning parameter for the magnitude of the disorder; and second, the disorder is weak~\cite{CainRomerSchreiber}. Therefore, we expect our system to be in the regime where a mobility edge exists, and eigenstates close to the center of the spectrum are delocalized, whereas states at the edge of the spectrum are localized~\cite{CainRomerSchreiber,  Biswas2000, Nikolic2001, Taraskin2002, Evangelou2003, Nishino_2008, Tadjine_2018}.

To probe the existence of mobility edges, we compute the generalized inverse partition ratio IPR$_q$~\cite{Evers_2008}, for several system sizes, focusing on $q=2$ and $\frac{1}{2}$ and the associated multifractal exponents: 
\begin{equation}
{\rm IPR}_q = \sum_i \lvert \psi_i \rvert^{2q}
\qquad 
D_q = \frac{1}{1-q} \frac{\ln ({\rm IPR}_q)}{\ln L^3} 
\, , 
\label{eq:IPR_Dq}
\end{equation}
where $L^3$ is the total number of sites on the cubic lattice, and $\psi_i$ is the spinon eigenstate wave function at site $i$. In the localized phase, both IPR$_{2,\frac{1}{2}}$ saturate to a constant value as a function of $L$, i.e., IPR$_{2,\frac{1}{2}} \sim \mathcal{O}(L^0)$, for $L \gg 1$. Instead, in the extended phase, we expect IPR$_q \sim L^{3(1-q)}$, for $L \gg 1$. In the following, we shall denote disorder-averaged quantities with an overline, e.g., $\overline{D_q}$. From Eq.~\eqref{eq:IPR_Dq}, one clearly sees that the multifractal exponent tends to $0$ when the sampled states are localized and tends to $1$ when the sampled states are delocalized, for both $q=\frac{1}{2}$ and $2$ in the thermodynamic limit $L \to \infty$.

\begin{figure}[ht!]
\centering
\includegraphics[width=0.85\columnwidth]{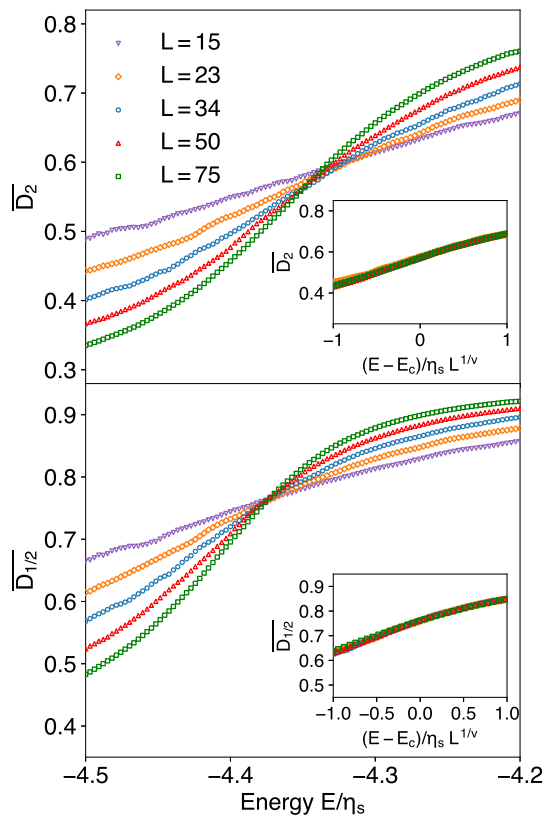}
\caption{\label{fig:mobility_edge}
Disorder-averaged multifractal coefficients $\overline{D_q}$ for $q=2$ (top panel) and $q=\frac{1}{2}$ (bottom panel) for various system sizes from $L=10$ to $75$, near the edge of the spectrum (in units of the spinon hopping). Our results suggest the existence of a mobility edge near $E_{c} \sim -4.35$. The insets show the scaling with system size for $\overline{D_q}$. The curves for various system sizes collapse onto the same curve when scaled with the empirical critical exponent $\nu=1.6$.
}
\end{figure}

The overall behavior of $\overline{D_q}$ across the energy spectrum is shown in Fig.~\ref{fig:mobility_edge}, for several $L$ as a function of $E$. The data were obtained using exact diagonalization, averaging over vison configurations~\footnote{Here, $1000$ vison configurations for $L=15$, $500$ for $L=23$, and $100$ vison configurations for larger system sizes.} and a small energy shell around the target energy~\footnote{The width of the energy shell has been chosen to be small enough for the change in the density of states to be negligible.}.

\begin{figure}
    \centering
    \includegraphics[width=0.42\textwidth]{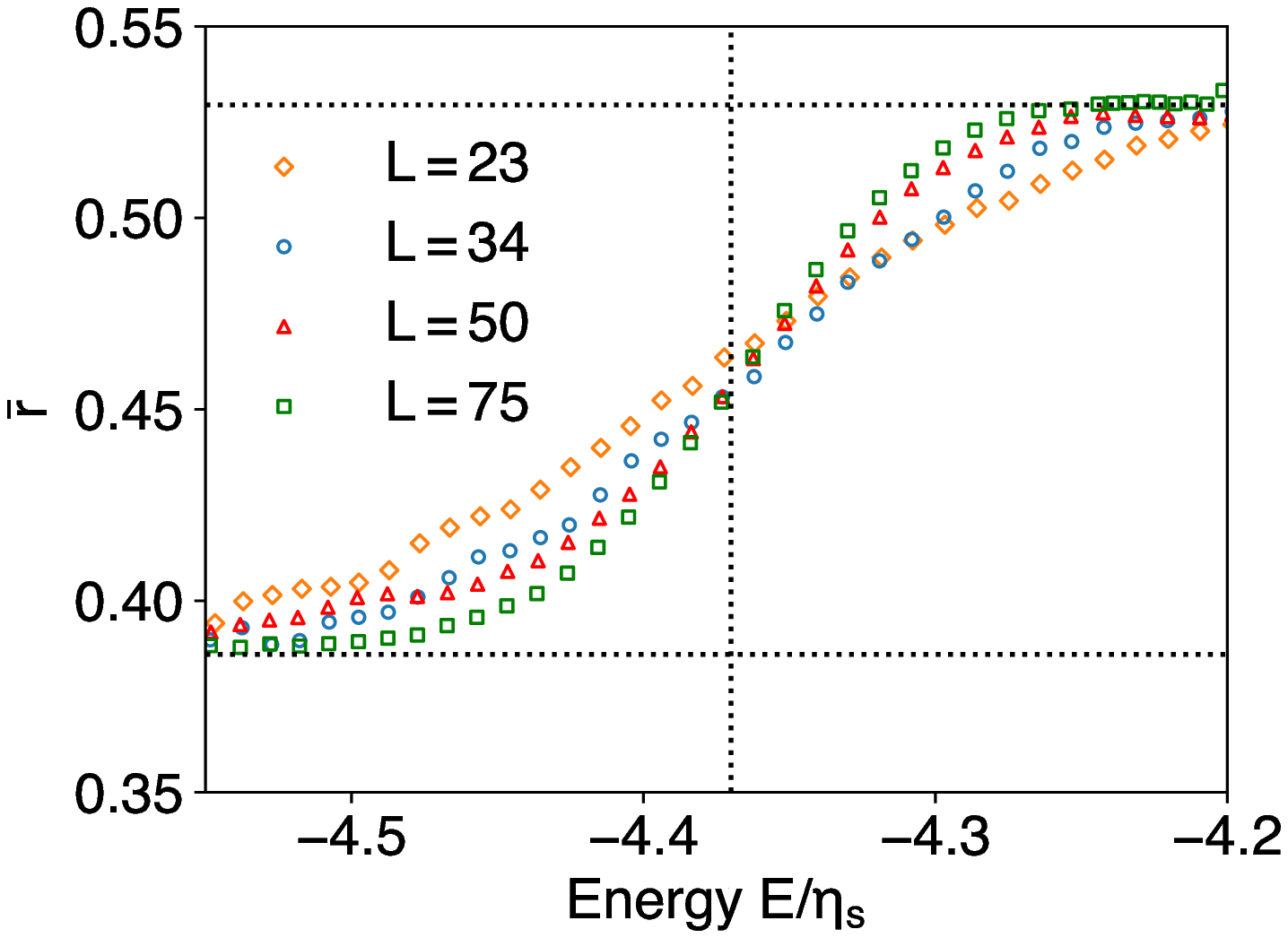}
    \\
    \includegraphics[width=0.42\textwidth]{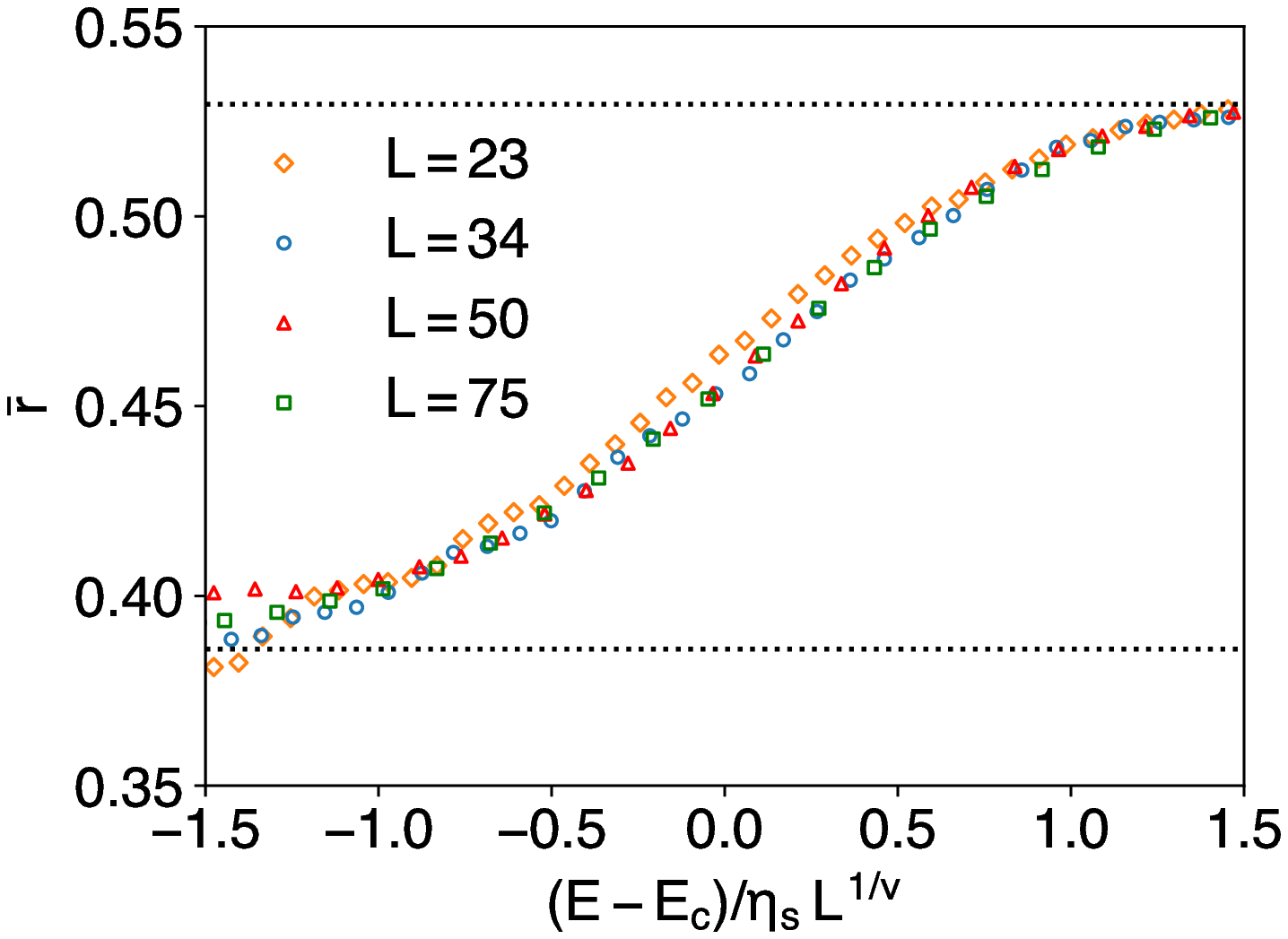}
    \caption{$r$ values window-averaged with window width $\Delta E=0.1$ (top panel), and the same data scaled using the empirical critical exponent $\nu=1.6$. The $r$ values transition from the Poisson distribution value (localized states, lower dotted line) to the Wigner value (extended states, upper dotted line), with a critical point at $E/\eta_s \sim -4.37$ (vertical dotted line). The results for various system sizes show a good collapse onto a single curve when scaled by system size near the mobility edge, analogously to Fig.~\ref{fig:mobility_edge}, suggesting the existence of a localization-delocalization transition.}
    \label{fig:rstat}
\end{figure}

We observe a general tendency toward delocalization in the middle of the spectrum and toward localization at the edge (akin to what was found for random bond disorder~\cite{CainRomerSchreiber, Biswas2000, Nikolic2001, Evangelou2003}), providing evidence of the existence of mobility edges. Upon closer inspection (see Fig.~\ref{fig:mobility_edge}), our results suggest the presence of a mobility edge around $E_{c} \sim -4.35$ (in units of the spinon hopping). Indeed, for $|E|>|E_{c}|$, $\overline{D_q}$ tends to zero with increasing $L$, and for $|E|<|E_{c}|$, we have $\overline{D_q} \rightarrow 1$. 
To better pin down the value of $E_{c}$, we perform finite-scaling analysis on $\overline{D_q}$ by collapsing the curves of $\overline{D_q}$ with different $L$. As shown in the insets in Fig.~\ref{fig:mobility_edge}, a good collapse is obtained using $E_{c}=-4.33$ for $q=2$ and $E_{c}=-4.37$ for $q=\frac{1}{2}$, with the critical exponent being $\nu = 1.6$ for both cases, in good agreement with the critical exponent of the 3D Anderson model with onsite disorder $\nu_{\mathrm{AT}} \approx 1.58$~\cite{Andersonscale1, Andersonscale2}, belonging to the orthogonal universality class. This is expected, given that our Hamiltonian is real and symmetric and therefore has time-reversal symmetry. In contrast, when one assumes random phases as off-diagonal disorder, time-reversal symmetry is broken, and it is known that the localization transition belongs to the unitary class~\cite{Kawarabayashi_98}. We note that this region, being close to the band edge, has a small DOS which is rapidly increasing as energy increases, as shown in Appendix~\ref{app:statistics}.
To further support the existence of mobility edges, we analyze the energy spectrum of the effective Hamiltonian $H_{\textrm{eff}}$ in Eq.~\eqref{eqn:MainHamiltonian}. We focus on the energy correlation between close-by energies, quantified by the so-called $r$-gap ratio~\cite{rstat, rstatdist}: 
\begin{equation}
    \label{rstatistic}
    r_n = \frac{\mathrm{min}(\delta_n,\delta_{n-1})}{\mathrm{max}(\delta_n,\delta_{n-1})}, \qquad \delta_n = E_n - E_{n-1}
    \, .	
\end{equation}
It is known that the mean value for $r$ should be $r_{\text{Poisson}} = 2 \log 2 - 1 = 0.386$ for localized states and $r_{\text{GOE}} = 0.5295$ for Gaussian orthogonal ensembles (GOEs, extended states)~\cite{rstat, rstatdist}. As shown in Fig.~\ref{fig:rstat}, at low energies, $r\approx r_{\text{Poisson}}$, and the system is localized, while for energies closer to the band center, $r_{\text{GOE}}\approx 0.52$, and the system is extended. Furthermore, in Fig.~\ref{fig:rstat} (bottom panel), we perform the same finite-sized scaling analysis for $r$, as in Fig.~\ref{fig:mobility_edge}. The analysis of the energy spectrum confirms $E_{c} \approx -4.37$ and the critical exponent $\nu\approx \nu_{\mathrm{AT}}$.

\begin{figure}[ht!]
\centering
\includegraphics[width=0.48\columnwidth]{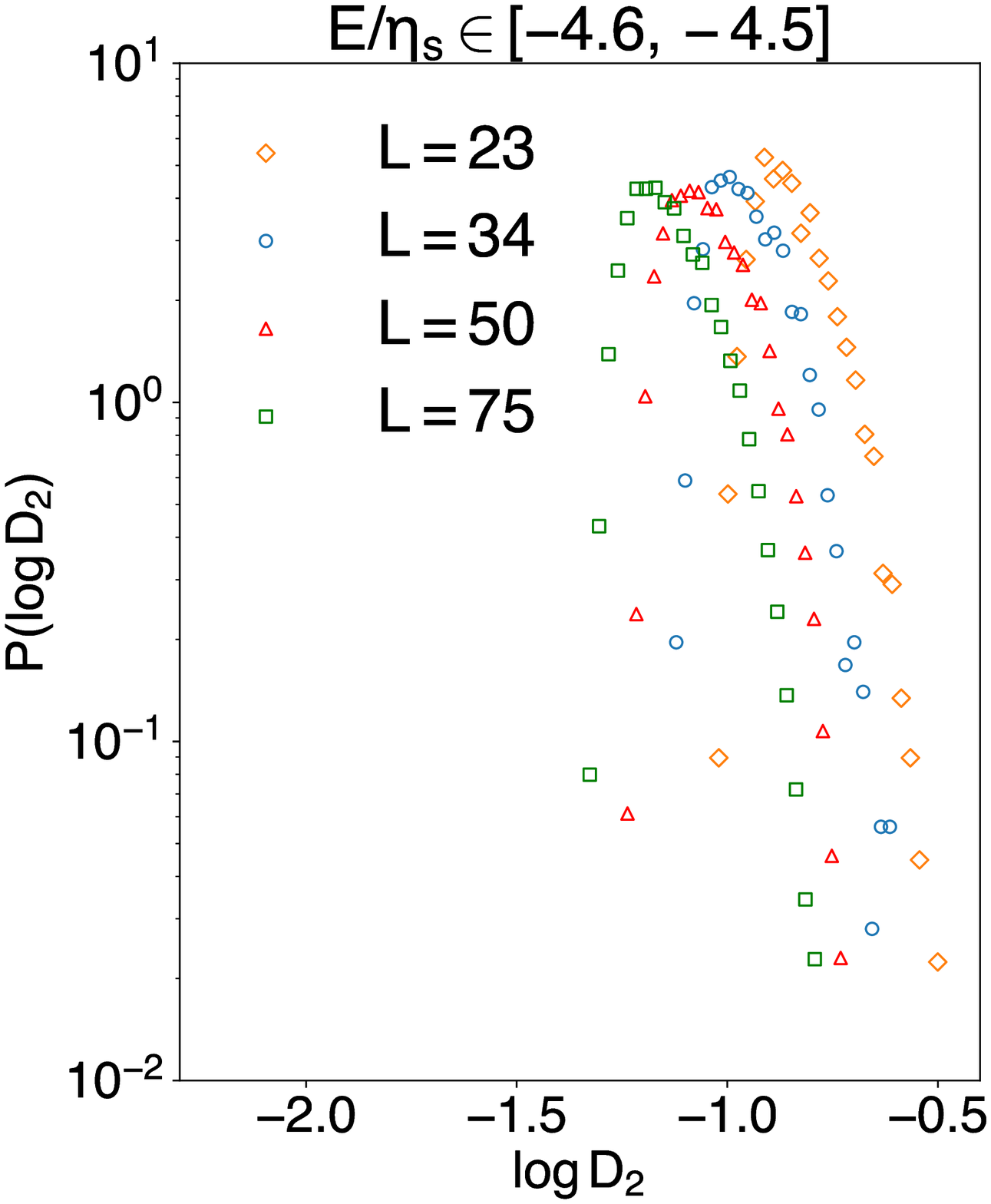}
\includegraphics[width=0.48\columnwidth]{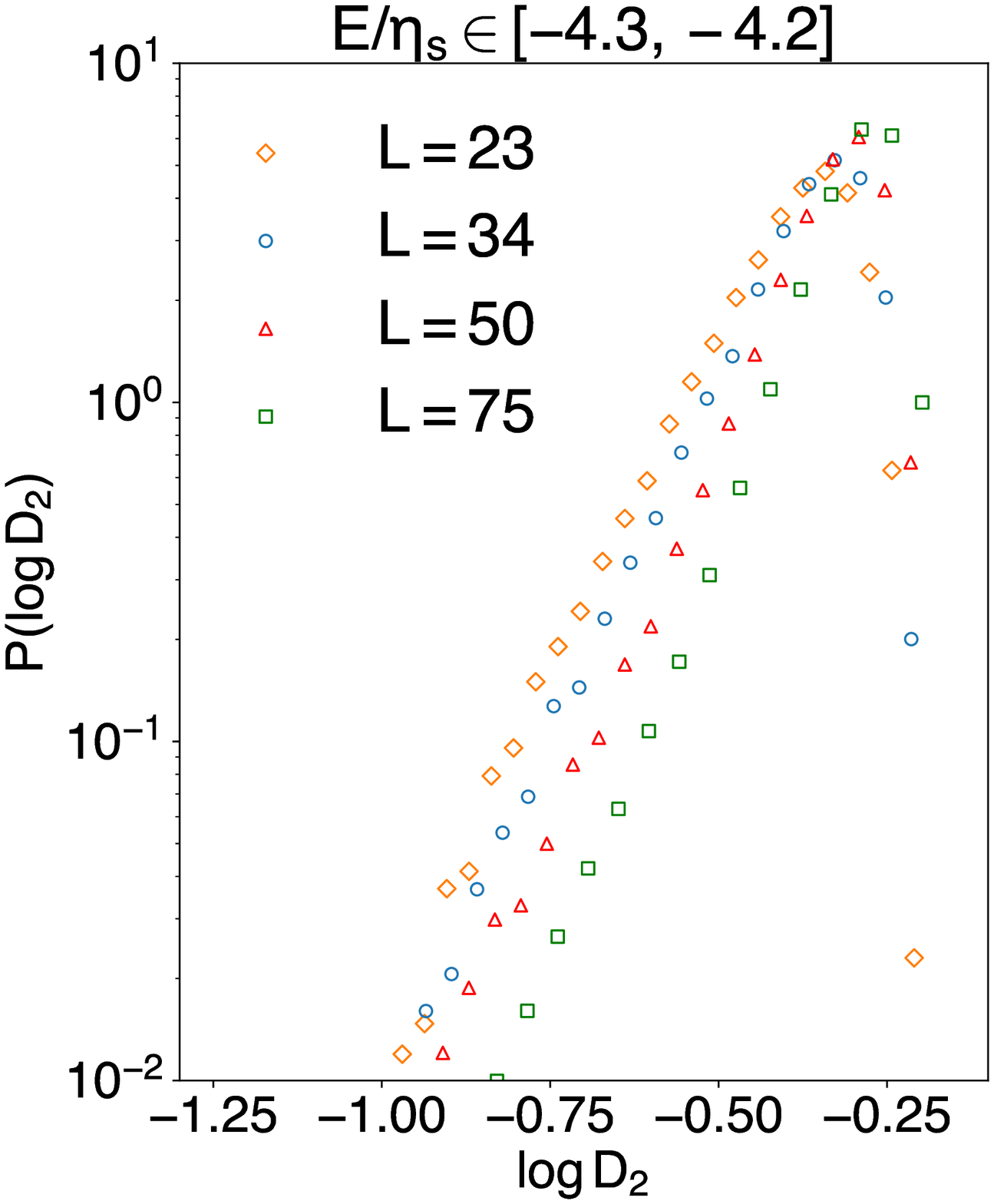}
\\
\includegraphics[width=0.48\columnwidth]{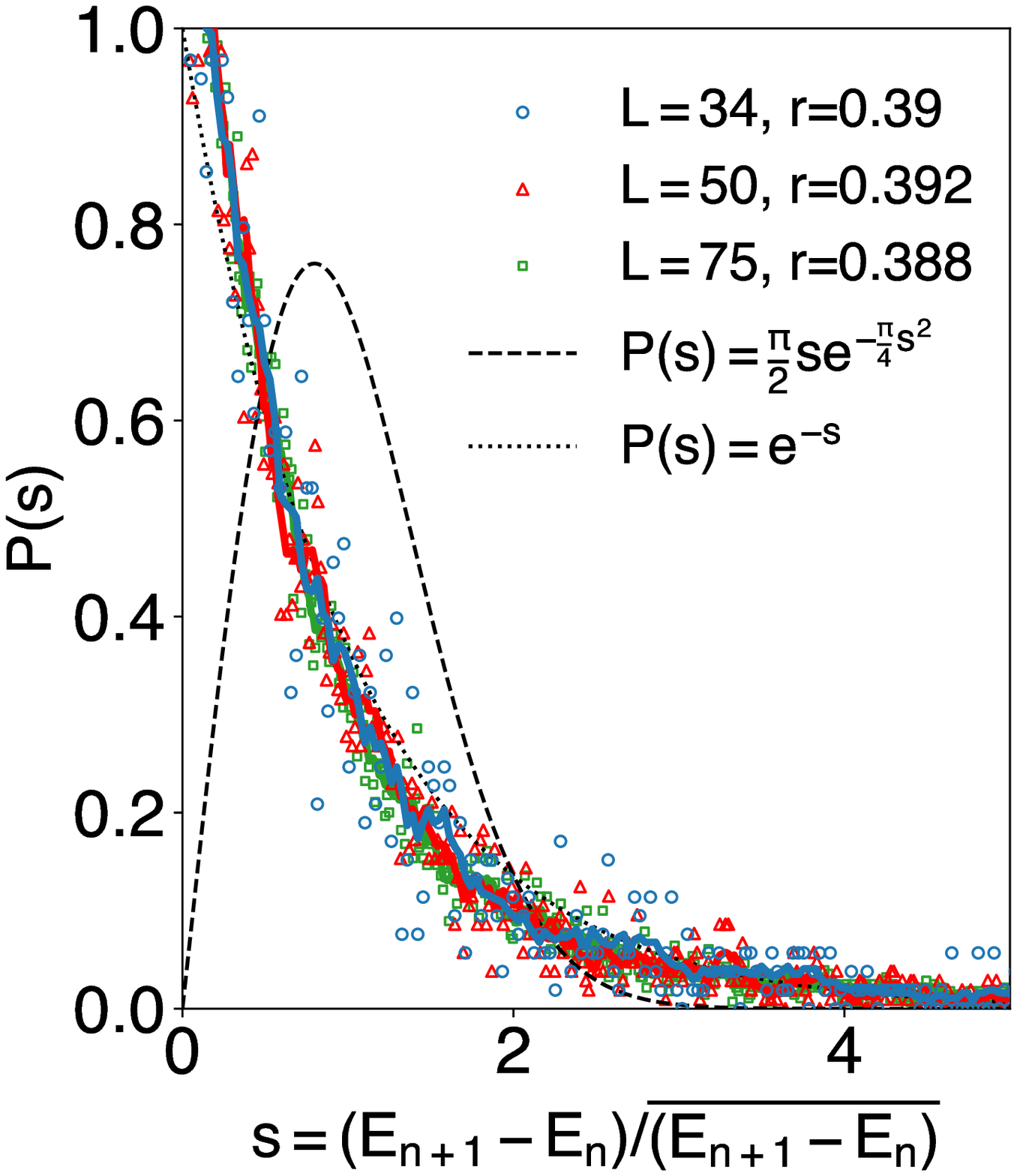}
\includegraphics[width=0.48\columnwidth]{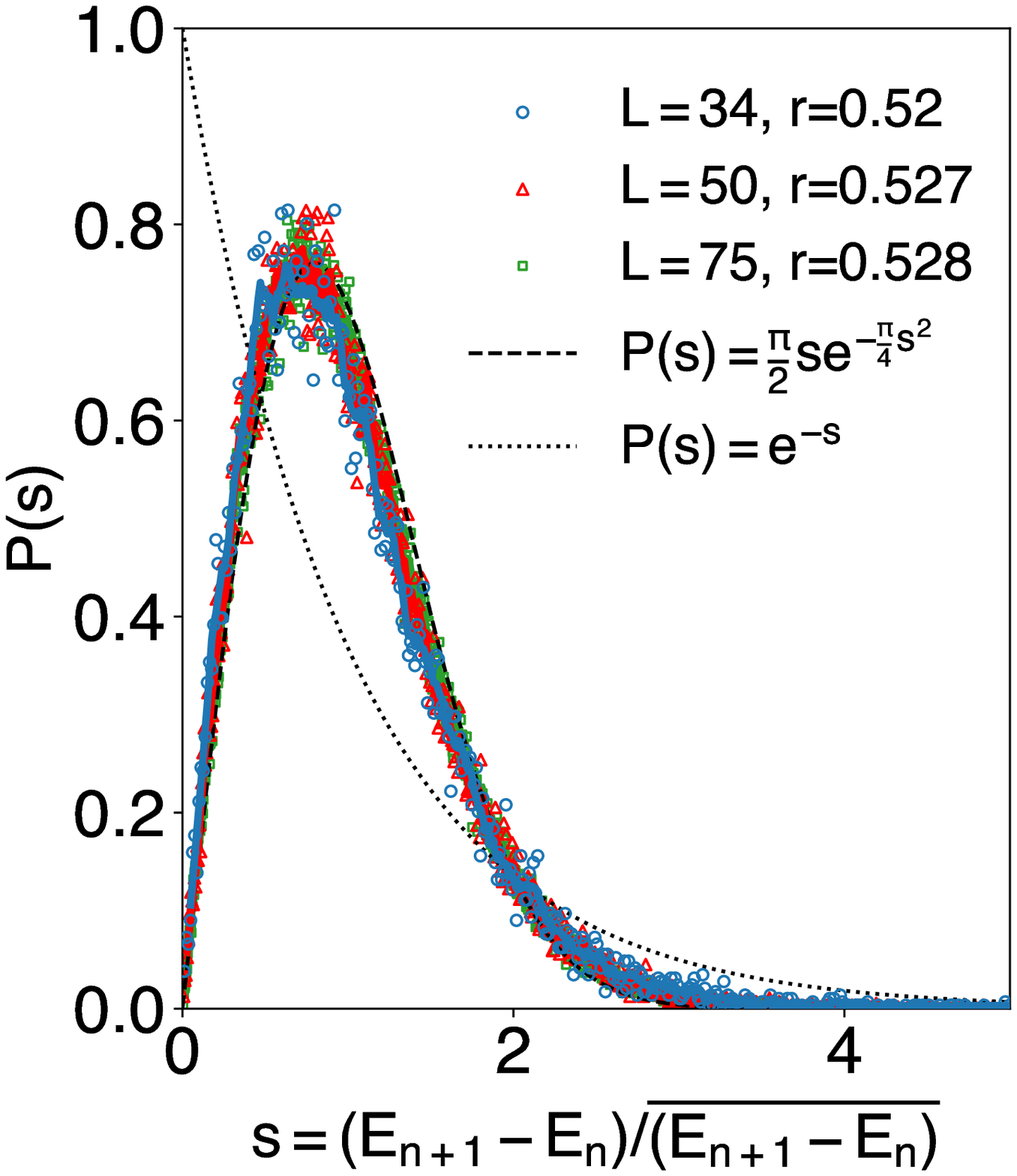}
\caption{\label{fig:IPRD2scale}
The probability distribution of the multifractal coefficient $D_2$ for localized ($E/\eta_s \in [-4.6, -4.5]$, left), and extended ($E/\eta_s \in [-4.3, -4.2]$, right) states, alongside the level statistics $P(s), s = (E_{n+1}-E_{n})/\overline{(E_{n+1}-E_n)}$ in the same energy windows. In the localized regime, we see a Poissonian distribution of the level statistics, while in the extended regime, we observe a Gaussian orthogonal ensemble (GOE) distribution (Wigner surmise). The $r$ value for each lattice size in the given energy window is also marked, showing once again that the states are localized (left panel) and extended (right panel). The dashed lines are the Poissonian distribution $P_{\text{Pois.}}(s) = \exp(-s)$ (left panel) and Wigner surmise $P_{\mathrm{GOE}}(s) = \frac{\pi s}{2}\exp(-\pi s^2/4)$ (right panel).
}
\end{figure}

Finally, for the sake of completeness, we inspect the probability distribution of $D_2$ and of the energy level spacing $s=\frac{E_{n+1}-E_n}{\overline{E_{n+1}-E_n}}$, for two energy windows, one in the localized phase and another in the extended phase (see Fig.~\ref{fig:IPRD2scale}). In the localized phase (Fig.~\ref{fig:IPRD2scale}, left panels), the probability distribution of $P(\log{D_2})$ shifts uniformly to lower values $\overline{D_q}\ll 1$ with increasing $L$, ruling out the existence of delocalized states. Correspondingly, the probability distribution of the level spacing is Poissonian, $P(s) \sim \exp(-s)$. In the extended phase (Fig.~\ref{fig:IPRD2scale}, right panels), both $P(\log{D_2})$ and $P(s)$ show full delocalization. 

One should note that, at finite temperature, thermal occupation of excited states and coupling to the bath can cause dephasing, which destroys localization. This is driven by variable-range hopping~\cite{Mott_69}, and the resulting conduction current (in 3D) can be expressed as $I = I_s \exp \left[ -\left(T_\mathrm{M} / T\right)^{1/4}\right]$, where $T_\mathrm{M}$ is the so-called Mott temperature. Then $I_s/I$ gives us an estimate of the timescale up to which the results discussed above may be expected to hold: $t_{\rm max} \sim \exp \left[ \left( T_\mathrm{M} / T \right)^{1/4}\right]$. In a localized state, $T_\mathrm{M} \propto [ \xi_{\mathrm{loc}} \rho(E) ]^{-1}$~\cite{Mott_69}, where $\rho(E) \sim 10^{-2}\eta_s^{-1}$ is the DOS in the localized phase and $\xi_{\mathrm{loc}} \sim 1$ is the localization length. Therefore, in comparison with the temperatures of interest $T/\eta_s < 10^{-2}$, the Mott temperature is relatively large $T_M/\eta_s \gtrsim 10^2$, and localization survives up to long times $t_{\rm max} \gtrsim 10^4$.
%
%
%

\section{\label{sec:propagation}
Spinon propagation
        }
In this section, we look at the case of a single spinon moving in a static random vison background (with probability $\frac{1}{2}$ of vison occupancy $n_p = 0,1$, or equivalently, density $\langle n_p \rangle = \frac{1}{2}$), which corresponds to the model introduced in Sec.~\ref{sec:model} where the energy cost of visons is set to zero (i.e., an infinite temperature regime for the visons)~\footnote{We remark that our choice of random vison background respects the emergent gauge structure of the system, discussed in Sec.~\ref{sec:model}.}. This regime is relevant to the short-time, short-distance regime where the visons are treated quasistatically with respect to the motion of the spinons. 

From the time dependence of the spinon wave function, we compute the root mean square displacement: 
\begin{equation}
\langle \mathbf{r}^2 (t) \rangle = \sum_i d^2(i,i_0) |\langle i|\exp(-i H_{\textrm{eff}}t ) |i_0\rangle |^2,
\end{equation}
where $d(i,i_0)$ is the Euclidean distance between the initial site $i_0$ and site $i$. 
To compute $\langle \mathbf{r}^2 (t) \rangle$, we employ fourth-order Trotterized~\cite{SuzukiTrotter} time evolution on a $L^3=149^3$ cubic lattice, with open boundary conditions. The initial state is chosen to be a spinon wave function localized on a single lattice site at the center of the lattice. 

\begin{figure}[ht!]
\centering
\includegraphics[width=0.85\linewidth]{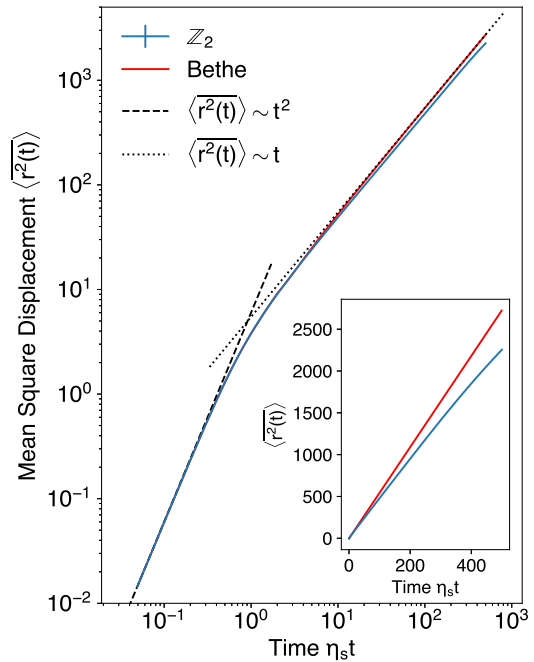}
\caption{\label{fig:propagation}\label{fig:Diffusion}
Numerical simulations of the spinon mean square displacement $\langle \mathbf{r}^2 (t) \rangle$ (blue line) compared with the analytical solution on the Bethe lattice (red line). We used fourth-order Suzuki-Trotter evolution on a $149^3$ lattice, averaged over $128$ different vison background realizations. The inset shows the same data in linear scale, to show more clearly the departure of the numerics from quantum diffusive behavior. 
}
\end{figure}

The results are shown in Fig.~\ref{fig:propagation}, where we see a crossover from short-range, short-time ballistic propagation $\langle \mathbf{r}^2 (t) \rangle \sim t^2$ to diffusivelike behavior $\langle \mathbf{r}^2 (t) \rangle \sim t$. This agrees with the previous results that our system belongs to the 3D universality class, which is characterized by a diffusive behavior within the extended phase~\cite{PRELOVSEK_1979, Prelov_1987,Ohtsuki_1997,Sierant_2020,Prelov_21}. 

The numerical results are contrasted with a self-retracing approximation that allows for an analytical solution on the Bethe lattice, as discussed in Ref.~\onlinecite{MainProp}. 
Details of the Bethe lattice calculation are given in Appendix~\ref{app:Bethe}. Here, we report and discuss only the main results, namely, the equation for the $2k$th moment $\langle \mathbf{r}^{2k}(t) \rangle$: 
\begin{equation}
\begin{split} \label{eqn:momentum}
\left\langle\mathbf{r}^{2 k}(t)\right\rangle &= \oiint \frac{d w_{1}}{2 \pi i} \frac{d w_{2}}{2 \pi i} \frac{\exp[i \eta_s t\left(w_{1}-w_{2}\right)]}{w_{1} w_{2}} T\left(\frac{1}{w_{1}}\right) T\left(\frac{1}{w_{2}}\right) \\
& \times \mathcal{R}_{2 k}\left[S\left(\frac{1}{w_{1}}\right) S\left(\frac{1}{w_{2}}\right)\right]
\, , 
\end{split}
\end{equation}
where $T(x)$ and $S(x)/x$ are generating functions for self-retracing paths from the root node (depth $0$) and nonroot node (depth $\geq 1$) for a Bethe lattice of coordination number $z$, and $\mathcal{R}_{2k}(x)$ is the generator for the $2k$th moment. The structure of $\mathcal{R}_{2k}(x)$ alongside $T(x)$ and $S(x)$ (see Appendix~\ref{app:Bethe}) implies that the pole signatures are the same as in 2D, with a line of double poles at $w_1 = u+i0^+, w_2 = u-i0^-$ for $u \in \mathbb{R}$ and $|u|<2 \sqrt{z-1}$. While there are subtle differences due to the additional lattice dimension (2D $\to$ 3D), the resulting identical pole signatures give the same power law dependence as in 2D; hence, we expect $\langle \mathbf{r}^2 \rangle \sim t^2$ at short times (ballistic propagation) and $\langle \mathbf{r}^2 \rangle \sim t$ at long times (quantum diffusion), with the crossover happening at $\eta_s t \sim 1$. This is indeed observed in Fig.~\ref{fig:Diffusion}.
In the self-retracing approximation, the 3D cubic lattice asymptotically exhibits the same diffusion coefficient as the 2D triangular lattice: 
\begin{equation}
D_{6}^{\mathrm{diff}} = \frac{3}{\pi}\left[3 \sqrt{5}-2 \ln \left(\frac{3+\sqrt{5}}{3-\sqrt{5}}\right)\right] 
\simeq 2.72968
\, . 
\label{eq:diffcoeff}
\end{equation}
In this paper, we use the superscript diff to unequivocally distinguish the diffusion
constant $D_k^{\mathrm{diff}}$, with $k$ being the coordination number of the lattice, from the multifractal exponent $D_q$. The full solution of the self-retracing approximation (Fig.~\ref{fig:Diffusion}) was obtained numerically, using the $\lvert w_1\rvert = \lvert w_2\rvert = 6$ torus (we are performing a double-loop integral; hence, the integration region is a torus, not a loop). This integration region encircles all the double-pole singularities. A numerically exact integral is achievable only up to $\eta_s t \sim O(1)$. Beyond that, one can demonstrate that the equations lead to asymptotic linear-in-time behavior, with the diffusion coefficient given above, Eq.~\eqref{eq:diffcoeff} [which is in excellent agreement with the behavior in the time window accessible by exact numerical integration of Eq.~\eqref{eqn:momentum}]. 

We observe good agreement between numerics and the Bethe lattice result up to $\eta_s t\sim 10^1$. We note that the subsequent departure from quantum diffusion, which is fitted by an anomalous exponent $t^{0.98}$ and unlikely to be significant, is much less pronounced than in 2D~\cite{MainProp}. 
This is in qualitative agreement with the 
results presented in Sec.~\ref{sec:mobility}, namely, the fact that the 3D system is generally less localized than its 2D counterpart.

One should bear in mind that, in general, the effective tight-binding picture derives from perturbation theory; hence, there is a time limit to the validity of our model. This is given by $t \lesssim \Delta_s/\eta_s^2$ when next-nearest neighbor hopping effects are expected to become relevant~\cite{MainProp}. After this timescale, the chirality of the system need not be obeyed, and the long-timescale behavior could deviate from typical expectations. 
%
%

\section{\label{sec:spinonvison}
Spinons and visons
        } 
Next, we look at the long time regime, where we adopt as in 2D~\cite{MainCorr} a Born-Oppenheimer perspective, with the spinons instantaneously in equilibrium on the timescales over which visons relax (via a stochastic MC process). Given the system sizes we can access, we limit ourselves again to the case of a single spinon in Eq.~\eqref{eqn:MainHamiltonian}. 

In our simulations, we use a cubic lattice of size $L^3 = 16^3$ with periodic boundary conditions. We initialize the system in a random vison background of density $\langle n_p \rangle = \frac{1}{2}$ (subject to the gauge conditions discussed in Sec.~\ref{sec:model}). We then run MC updates using a Metropolis algorithm as follows:
(i) We pick one bond $i$ at random. There are four plaquettes which contain $i$. We propose an update which flips the vison number of those four plaquette $n_p \to 1-n_p$. 
(ii) A new effective spinon Hamiltonian $H'$, following Eq.~\eqref{eqn:MainHamiltonian}, is constructed by changing $A_{ss'}$ to $A_{ss'}+\pi$ for the bond $i \equiv ss'$ that has been selected at point (i). 
(iii) We accept the proposed change according to the Metropolis probability $\textrm{min}[1, \textrm{tr} \: \exp(-\beta H')/\exp(-\beta H)]$, where $H$ is the spinon Hamiltonian before the update, and $\beta = 1/T$ is the inverse temperature (working in energy units where the Boltzmann constant is $k_\mathrm{B} = 1$). 

We are interested in a regime where the vison energy cost $\Delta_v$ does not play a significant role and we set it to zero for convenience~\cite{MainCorr}.
Moreover, at the temperatures of interest in this paper ($T/\eta_s \leq 0.01$), we find that it is sufficient to consider only the six lowest eigenstates when computing the traces above, and the contribution of higher excited states can be safely neglected. 
This simplification greatly enhances the performance of our code.

\begin{figure}[ht!]
    \centering
    \subfloat{
		\includegraphics[width=0.47\linewidth]{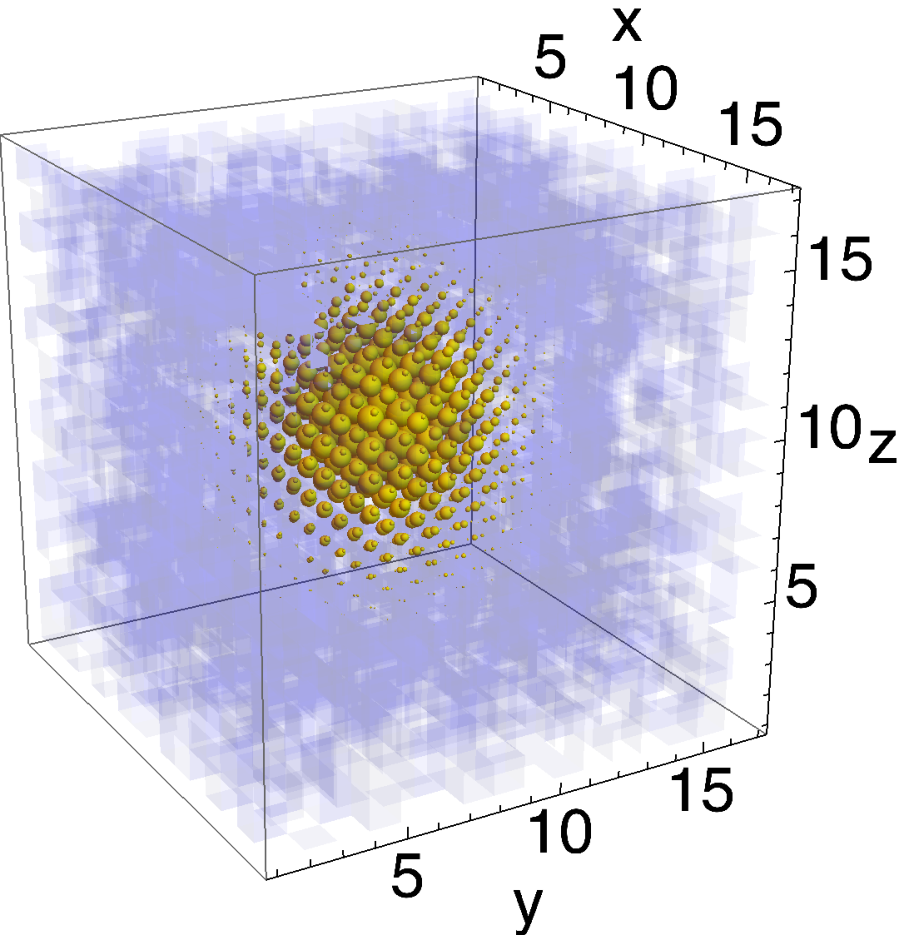}
	}
	\subfloat{
        \raisebox{0.25\height}{\includegraphics[width=0.47\linewidth]{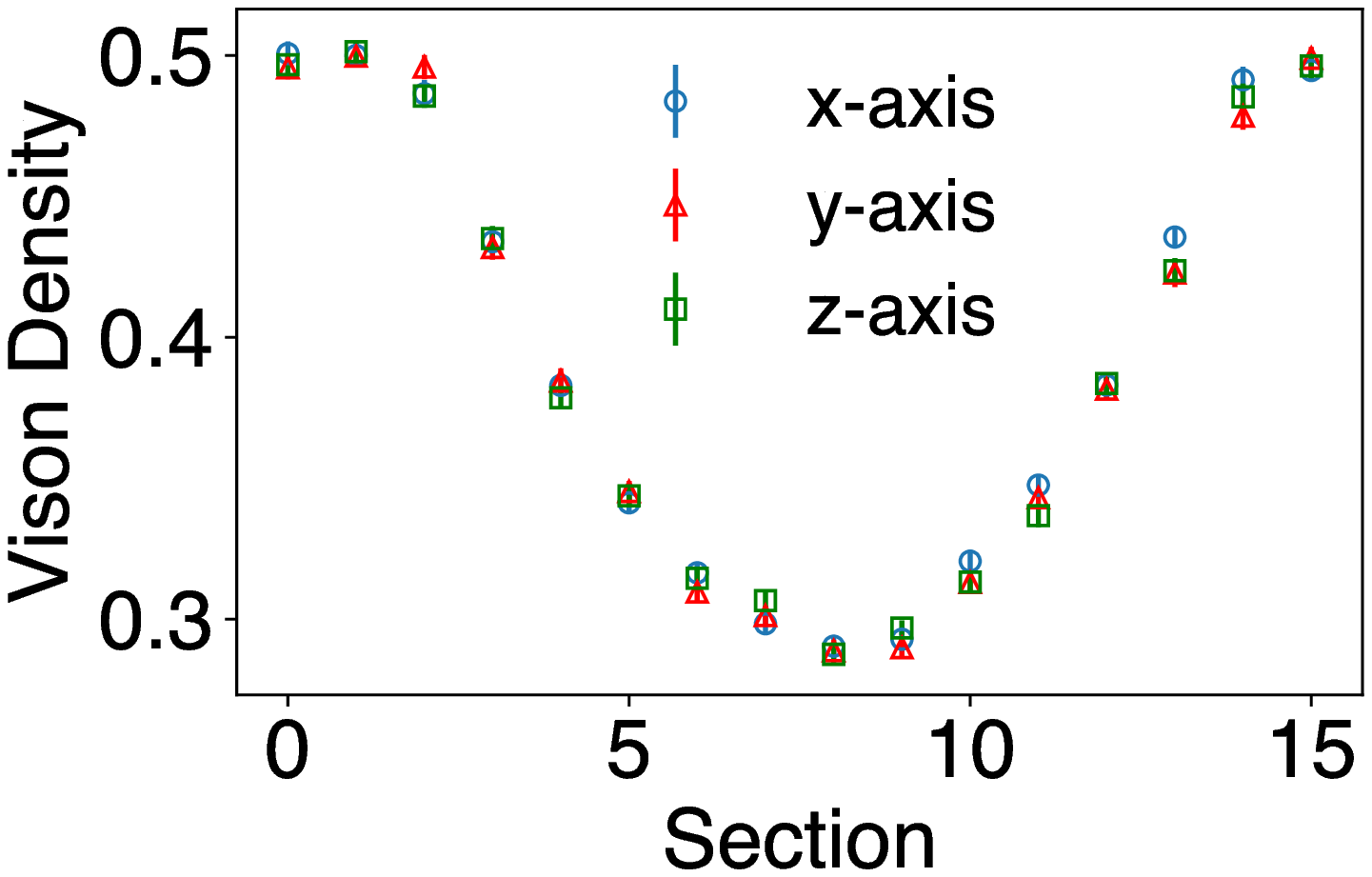}}
    }
    \vskip\baselineskip
    \centering
    \subfloat{
    \includegraphics[width=0.95\linewidth]{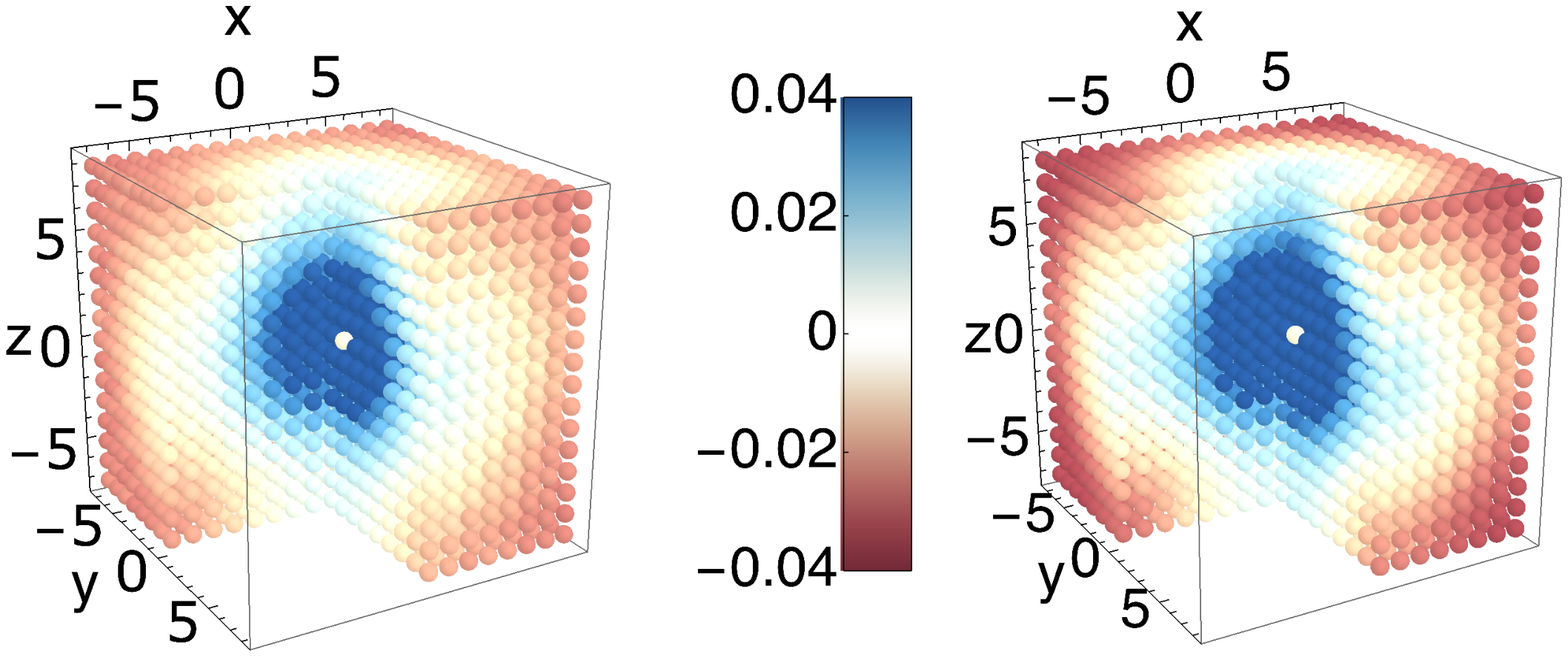}
    }
    \caption{\label{fig:SpinonVison}
    Vison configuration (blue squares) overlapped with the spinon probability density (yellow spheres, radius proportional to probability) at $T_1/\eta_s=1.0 \times 10^{-4}$ (top left panel).
    The depletion region has been shifted to the center of the system for visualization convenience, taking advantage of the periodic boundary conditions. 
    The top right panel shows the vison density for each layer of the $L^3=16^3$ lattice, as a function of layer index for each of the three coordinate axes; this illustrates quantitatively the presence of a spherical vison depletion region. 
    The bottom panels are the connected vison correlators $C_\rho(r)$, Eq.~\eqref{eq:conncorr}, at the same temperature: left panel from Monte Carlo simulations; right panel from a phenomenological model which assumes a vison-depleted sphere surrounded by a randomly half-filled vison background.}
\end{figure}

Despite the presence of a mobility edge and delocalized states in the spectrum (see Sec.~\ref{sec:mobility}), we clearly observe the formation of a vison depletion region around the spinon wave function and a rapid decay of the latter outside the depleted region. 
Correspondingly, most of the results appear to be like the 2D case~\cite{MainCorr}. At low temperature, we observe the appearance of a spherical region where the vison density vanishes, and the spinon wave function is finite almost exclusively inside this region. Outside, the spinon density vanishes, and the vison density rapidly approaches the random uncorrelated value $\frac{1}{2}$. 
This is illustrated in Fig.~\ref{fig:SpinonVison}, where a specific instance is shown of a vison configuration and corresponding spinon density. 
(The spinon density was approximated here for simplicity by the square modulus of the ground state spinon wave function since it carries most of the Boltzmann weight; indeed, we find that $\Delta_{01}/\eta_s ~ O(10^{-1}) \gg T_1/\eta_s$, where $\Delta_{01}$ is the spinon energy gap above the ground state.) 
We further confirm this behavior statistically by computing the vison connected correlation function: 
\begin{equation}
C_\rho(s, s') 
= 
\frac{1}{3} \sum_{\alpha = x,y,z} \left[ 
\langle n_{p_\alpha} n_{p^{'}_\alpha} \rangle 
- 
\langle n_{p_\alpha} \rangle 
\langle n_{p^{'}_\alpha} \rangle 
\right] 
\, , 
\label{eq:conncorr}
\end{equation}
also shown in Fig.~\ref{fig:SpinonVison}. 
Here, $p_\alpha$ labels the plaquette adjacent to site $s$ that lies perpendicular to the direction $\alpha$; similarly, $p^{'}_\alpha$ labels the plaquette adjacent to site $s'$ that lies perpendicular to the direction $\alpha$ (where we have uniquely labeled all plaquettes on the cubic lattice by associating one plaquette for each of the three directions to one given site of the lattice). 
The angular brackets represent the thermodynamic equilibrium average computed in our MC simulations. 
%
%

\subsection{\label{sec:free_energy}
Effective free energy
           }
As for the 2D case~\cite{MainCorr}, the behavior of the system can be understood using a simple effective free energy model, where the spinon is confined to a spherical region of radius $\xi$ devoid of visons, and the visons outside the region take on a random configuration of density $\langle n_p \rangle = \frac{1}{2}$. 
In this limit, the spinon energy levels are known analytically [we consider for simplicity only the ground state $E(\xi) = \pi^2 \eta_s / \xi^2$], and the visons contribute only entropically to the free energy of the system. 
To account in part for the decay of the spinon wave function into the region where visons are present (due to a short but finite localization length), an additional temperature-independent penetration length scale $\xi_0$ is introduced: $E(\xi) \to E(\xi+\xi_0)$. 
The effective free energy $F(\xi)$ can be written as
\begin{equation}
    \label{eqn:free_en_spinon}
    F(\xi) = 
    \frac{\pi^2 \eta_s}{(\xi+\xi_0)^2} 
    + 
    \frac{4}{3} \pi {\xi}^{3}T 
    \mathrm{ln}\,[1+\exp(-\beta {{{\Delta }}}_{\text{v}})]
    \, . 
\end{equation}
Minimization with respect to $\xi$ gives the saddle-point equilibrium value $\xi_*$, dependent on $T$ and $\xi_0$:
\begin{equation} \label{eqn:SaddleExact}
    \left[1 + \left(\frac{\xi_0}{\xi_*}\right)^3 \right] \xi_*^5 = \frac{\pi}{2 \ln 2} \frac{\eta_s}{T}
    \, , 
\end{equation}
where we set for simplicity $\Delta_v = 0$. This implicit equation can be solved numerically, and in the $\xi_* \gg \xi_0$ regime, it gives the scaling behavior $\xi_* \sim T^{-1/5}$, in contrast to the 2D scaling $\xi \sim T^{-1/4}$~\cite{MainCorr}. 

To account for thermal fluctuations, one can compute instead the equilibrium value of $\xi$ using the formula: 
\begin{equation} \label{eqn:expectvalue}
    \langle \xi \rangle =\frac{1}{Z}\int_{0}^{R}{d}\xi \ \xi \exp[-\beta F(\xi )]
    \, ,
    \quad 
    Z=\int_{0}^{R}{d}\xi \ \exp[-\beta F(\xi)]
    \, .
\end{equation}
Here, $R$ is a cutoff that captures the finite-sized effect in the MC simulations. 
At the temperatures of interest in this paper, we observe that the Boltzmann distribution for $\xi$ is in fact highly peaked, and the saddle-point value $\xi_*$ is a very good approximation for $\langle \xi \rangle$. 

\begin{figure}
    \centering
    \includegraphics[width=0.48\textwidth]{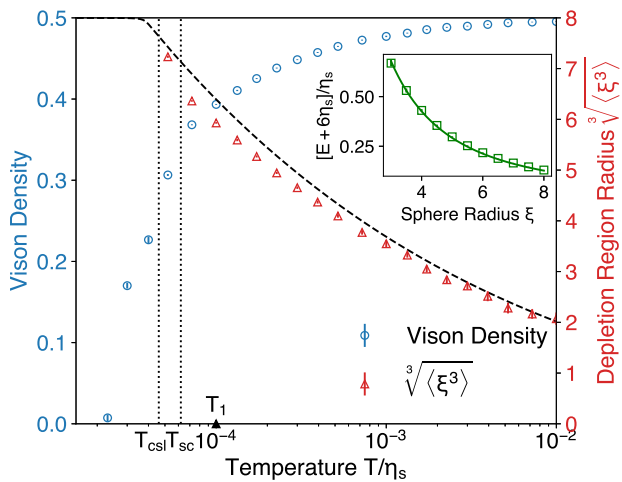}
    \caption{\label{fig:MainDiagram}
    Dependence of the average vison density $\langle n_p \rangle$ per plaquette (blue circles, left vertical scale) and of the radius of the depleted region $\sqrt[3]{\xi^3}$ (red triangles, right vertical scale) as a function of temperature. 
    The dashed black line is the analytical result from the effective model in the main text, Eq.~\eqref{eqn:free_en_spinon}.
    The vertical dotted lines indicate the temperatures at which finite-sized instabilities occur in the shape of the depleted region, from spherical to cylindrical $T_{\mathrm{sc}}$ and from cylindrical to planar slab $T_{\mathrm{csl}}$ as discussed in Appendix~\ref{app:strip}. $T_1$ is the temperature used in Fig.~\ref{fig:SpinonVison}, as a reference. The simulations were performed on an $L^3=16^3$ lattice with periodic boundary conditions, averaged over $50$ independent runs. 
    The inset shows fitting the ground state energy of a spinon in an artificial vison background generated according to Eq.~\eqref{eq:toyxi0} on a $L^3=20^3$ lattice, as a function of $\xi$ and averaged over $500$ realizations (squares), to the energy $E(\xi) = \pi^2 \eta_s / (\xi + \xi_0)^2$ discussed in the main text (solid line) to obtain $\xi_0=1.814(0)$. The vertical axis has been shifted and rescaled for convenience.}
\end{figure}

To determine $\xi_0$, we compare the analytical result $E(\xi) = \pi^2 \eta_s / (\xi + \xi_0)^2$ with simulations of a toy system where the visons are artificially set to random configurations of probability:
\begin{equation}
p({n}_{p}) = \left\{
\begin{array}{ll}0,&\,\text{if}\,| {{\bf{r}}}_{p}|\, <\, \xi 
\, ,
\\ 
\frac{1}{2},&\,\text{otherwise} 
\, .
\end{array}
\right.
\label{eq:toyxi0}
\end{equation}
The ground state energy of the spinon averaged over $100$ independent vison configurations is shown in the inset of Fig.~\ref{fig:MainDiagram}, demonstrating good agreement for $\xi_0 = 1.814(0)$.

Equipped with the best fit value of $\xi_0$, we can then compare the equilibrium behavior predicted by the effective model in Eqs.~(\ref{eqn:free_en_spinon} and (\ref{eqn:expectvalue}) with the MC simulations of the spinon and vison system introduced at the start of this section. 
This is done in Fig.~\ref{fig:MainDiagram} for the vison density and the radius of the vison depletion region. 
We show $22$ temperature points from $\textrm{log}_{10} T/\eta_s = -4.625$ (where the vison density reaches the $1/{\rm volume}$ limit for our finite system size) to $\textrm{log}_{10} T/\eta_s = -2$, each averaged over $50$ independent cooling histories. 
The effective free energy model appears to capture the behavior of the system decently well. 

As temperature is lowered, the size of the vison depletion region increases. In a finite-sized system, it eventually percolates, and in doing so, we observe that it deforms first into a cylinder and then into a planar slab. This is a finite-sized effect that is largely outside the scope of this paper; however, for completeness, we present some results about it in Appendix~\ref{app:strip}. The temperatures at which these two instabilities occur, according to our simulations, are $T_{\mathrm{sc}}/\eta_s = 6.1 \times 10^{-5}$ (sphere to cylinder) and $T_{\mathrm{csl}}/\eta_s = 4.5 \times 10^{-5}$ (cylinder to planar slab) and are indicated for reference by vertical dotted lines in Fig.~\ref{fig:MainDiagram}. 
%
%

\subsection{\label{sec:interactions}
Mutual statistics vs interactions 
           }
The behavior discussed so far is entirely driven by the mutual statistics between the noninteracting quasiparticles. The spinon localization that underpins it, however, could in principle be driven also by possible spinon-vison interactions. Therefore, it is important to contrast the two cases explicitly. 

For concreteness, we consider the case of contact interactions of strength $U = \eta_s/4$ added to the spinon tight-binding model above: 
\begin{equation}
    H_{\textrm{int}} = H_{\textrm{eff}} + U \sum_s \left(\frac{1}{12} \sum_{s \in p} n_p \right) b_s^\dagger b_s
    \, . 
\end{equation}
Here, $H_{\textrm{eff}}$ is the effective Hamiltonian in Eq.~\eqref{eqn:MainHamiltonian}, and $s \in p$ indicates that the sum runs over the plaquettes $p$ which are adjacent to the spinon lattice site $s$.

We then proceed to compare the behavior of the new system with ($A_{ss'} \neq 0$, $\pi$ flux) and without ($A_{ss'}=0$, zero flux) mutual statistics between the spinons and the visons. In the latter case, $H_{\rm int}$ describes a 3D tight-binding system with uniform hopping strength $\eta_s$ and diagonal disordered potential, which is generally expected to contribute to the spinon localization.

\begin{figure}
    \centering
    \subfloat{
    \includegraphics[width=0.95\linewidth]{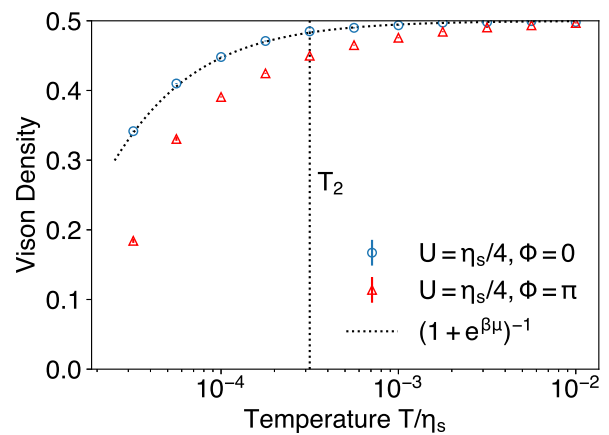}
    }
    \vskip\baselineskip
    \subfloat{
    \includegraphics[width=0.95\linewidth]{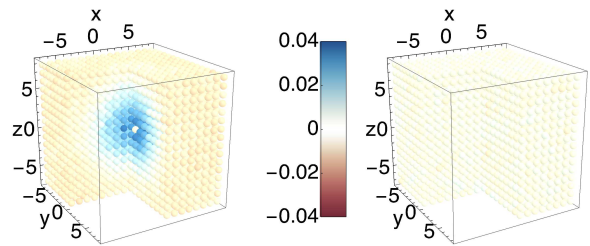}
    }
    \caption{\label{fig:fluxnoflux}
    Vison density as a function of temperature in the presence of spinon-vison contact interactions of strength $U=\eta_s/4$, with (red triangles) and without (blue circles) a mutual statistical angle ($\phi=\pi$ vs $\phi = 0$). The behavior when there are no mutual statistics between spinons and visons is well captured by a trivial change in the overall chemical potential for the visons ($\mu = 2.134 \times 10^{-5} \eta_s$ gives a good fit), in stark contrast with the dependence observed in the original model.
    Correspondingly, the vison density decreases in an essentially uniform way across the system, instead of forming a spherical depleted region. This is illustrated by the behavior of the connected correlator $C_\rho (r)$ in the bottom panels, for the case with (left) and without (right) mutual statistics.
    The correlators are computed at $T_2/\eta_s = 3.16 \times 10^{-4}$. Each data point is averaged over $32$ separate histories.}
\end{figure}

In the regime of interest to us ($U < \eta_s$, as we expect any interactions --- if at all present --- to be weak perturbations of the original model), localization due to diagonal disorder occurs over long length scales. This is in stark contrast with the very pronounced localization effects due to semionic statistics (equivalently, $\pi$ fluxes). 
Correspondingly, in our simulations, we do not observe a well-defined vison depletion region in the purely interacting case; the vison density is suppressed due to the presence of spinons in the system, but this occurs in a uniform way that is in fact well captured by a simple modeling with a tunable uniform chemical potential. Figure~\ref{fig:fluxnoflux} illustrates the dependence of the vison density as a function of temperature for the two cases at hand. 
Once again, as it was the case in 2D~\cite{MainProp, MainCorr}, we find that the effects of mutual statistics are robust and more pronounced than (reasonable) spinon-vison interactions, suggesting that evidence in candidate $\mathbb{Z}_2$ spin liquid systems of spinon localization in vison-depleted regions would be a signature of anyonic statistics, fractionalization and of a topologically ordered QSL. 
%
%

\section{\label{sec:out-of-equilibrium}
Response out-of-equilibrium
        } 
\begin{figure}
    \centering
    \includegraphics[width=\linewidth]{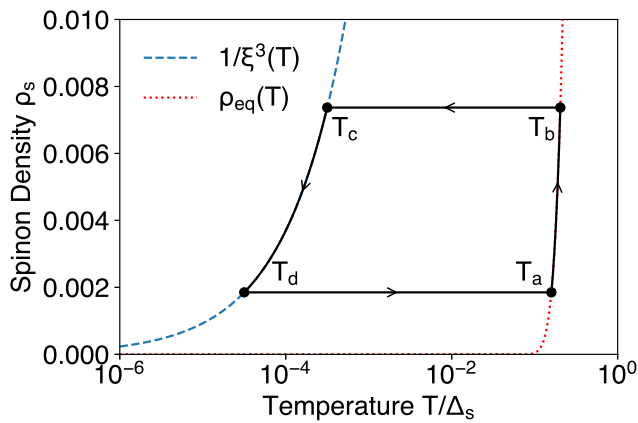}
    \caption{\label{fig:rho_hist}
    Schematic illustration of the spinon density out of equilibrium. The system is initially at $T_b$ (see main text). Upon lowering the temperature, the spinon density $\rho_s$ is unable to relax due to the slow diffusion of the spinons dressed by their vison depletion regions. The latter, however, grow in size as temperature is lowered and eventually come into contact with one another at $T_c$, when the correlation radius $\xi^*$ becomes comparable with the spinon separation distance $\sim \rho_s^{-1/3}$. This produces a kinematic locking of the spinon density to the $1/(\xi^*)^3 \sim T^{3/5}$ curve. When the thermal sweep is reversed at $T_d$ and the temperature is increased, once again the spinon density enters a plateau as annihilation events stop, until thermal equilibrium is restored $\rho_{\rm eq} \sim \exp(-\Delta_s / T)$.}
\end{figure}

Finally, we briefly discuss some of the effects that the phenomenology encountered in our system has on its behavior out of equilibrium. Following the 2D case in Ref.~\onlinecite{MainCorr}, we consider sweeps where the temperature is varied at a finite rate, starting from an equilibrium state at $T_b$ where the spinon density $\rho_s \sim \exp(-\beta \Delta_s)$ is much smaller than $1/\xi^3$ (as schematically illustrated in Fig.~\ref{fig:rho_hist}). 

Upon lowering $T$, the system ought to lower its spinon density to remain in equilibrium. Given the nature of the initial state, spinon annihilation processes (which can only occur in pairs) require them to travel cooperatively with their own vison depletion region for large distances before they meet another spinon. Like the 2D case, this can be expected to be a slow process, and if $T$ is lowered at a sufficiently fast rate, the system then falls strongly out of equilibrium at some constant spinon density. 

While the spinon motion is expected to be slow (cooperative), the size of the vison depletion region can easily adjust to remain in thermal equilibrium (driven by vison creation/annihilation events at the edge of the region). As temperature is lowered, one therefore expects the depletion regions to grow around each spinon --- thus ultimately and necessarily curtailing the constant spinon density regime. Once the depletion regions, and therefore the corresponding spinon wave functions, become large enough to overlap, spinon annihilation processes resume. In the effective free energy picture, this occurs when 
\begin{equation}
\rho_0 \, \xi_*^3 \sim 1
\:
\leftrightarrow 
\:
T \sim \eta_s \, \rho_0^{5/3}
\, , 
\label{eq:kinematic_limit}
\end{equation}
where $\rho_0$ is the initial spinon density. The right-hand side of the equation is obtained using the scaling relation $\xi_* \sim T^{-1/5}$ derived in the large $\xi_*$ limit in Sec.~\ref{sec:free_energy}. 

As we continue to lower the temperature, the spinon density enters a new kinematically locked regime where it decreases at the rate set by the condition that the typical depletion region diameter equals the mean separation between spinons (see Fig.~\ref{fig:rho_hist}). 
Using again the left-hand side of the expression in Eq.~\eqref{eq:kinematic_limit}, one finds the kinematically locked spinon density: 
\begin{equation}
\rho_{\rm kl} \sim \xi_*^{-3}
\sim T^{3/5} 
\, . 
\end{equation}
This differs quantitatively from the 2D case~\cite{MainCorr}, where it was found that $\rho_{\rm kl}(T) \sim T^{1/2}$.

When the thermal sweep direction is reversed, the vison depletion regions contract in size as temperature rises, and spinon annihilation stops. This generates a new plateau in the spinon density at some $\rho_s$ much greater than the thermal equilibrium value. This plateau lasts until the temperature becomes high enough to restore equilibrium $\rho_s \sim \exp(-\Delta_s/T)$, at which point the spinon density resumes increasing following the adiabatic curve, completing the hysteresis loop (as illustrated in Fig.~\ref{fig:rho_hist}).
%

\section{\label{sec:conclusions}
Discussion and conclusions
        } 
In this paper, we studied the behavior of a 3D $\mathbb{Z}_2$ lattice gauge theory (toric code)~\cite{3DToric, Toric3DCastelnovoChamon}, in the finite-temperature regime where the spinon excitations are sparse and can hop coherently across the system, whereas the vison excitations are quasistatic and stochastic. The latter therefore act as $\mathbb{Z}_2$ ($\pi$) fluxes for the former via their mutual semionic statistics. 
This model exhibits an instance of Anderson localization \cite{Anderson_58} driven by anyonic statistics, where both the noninteracting tight-binding particles as well as the disorder are borne out of the same spin degrees of freedom in an otherwise disorder-free topological QSL. 

We investigated the localization properties of the system and found that it exhibits a mobility edge, which is pushed toward the end of the spectrum. We determined the corresponding transition energy using various standard techniques such as level statistics~\cite{semipoisson, Evangelou2003}, $r$-statistics \cite{rstat, rstatdist}, and IPR/multifractal coefficients \cite{Evers_2008}. The vicinity of the mobility edge to the tail end of the spectrum leaves only a small proportion of states in the localized region at the extreme ends of the spectrum. By using finite-sized scaling, we provide evidence that the system belongs to the 3D Anderson universality class \cite{Andersonscale1, Andersonscale2}, and therefore, states inside the mobility edges are fully extended and delocalized. 

Furthermore, we inspect the out-of-equilibrium dynamics. As expected, we found that, at long times, the system shows diffusion as the 3D Anderson model with onsite disorder~\cite{PRELOVSEK_1979, Prelov_1987,Ohtsuki_1997,Sierant_2020,Prelov_21}. This result should be compared with the 2D case, where a strong subdiffusive-like behavior was observed~\cite{MainProp}. Finally, we support our numerical evidence with an analytical computation using a Bethe lattice approximation. 

Interestingly, once the visons are allowed to relax stochastically in the presence of a spinon, the behavior observed in 3D remains remarkably close to the one encountered in 2D~\cite{MainProp}, despite the presence of a mobility edge and delocalized states. We observe that visons are expelled from the high spinon density area, forming a vison-depleted region determined by the balance between spinon kinetic energy and vison configurational entropy. The effective energy stipulates that the size of the patches have a typical radius that scales with temperature as $T^{-1/5}$. The close resemblance between 2D and 3D behavior despite a mobility edge is probably due to the fact that low-energy states close to the lower edge of the band dominate in importance in the thermodynamic behavior that leads to the appearance of the vison depletion region –-- hence, the localized behavior dominates. We note, however, a longer penetration depth of the spinons into the random vison background compared with 2D, hinting at weaker localization in 3D.

As in the 2D case, such nontrivial interactions between spinons and visons are driven by mutual statistics. We have indeed shown that the spinon-vison coupling by itself is not enough to induce clearly visible vison-depleted regions. In this paper, we ignored possible interaction terms between the same quasiparticle species, which may however be present in real materials. While the topological spin liquid behavior is protected against sufficiently weak interactions by the presence of a finite gap, the localization properties are subject to change in what would then be a many-body localized system.
Nonetheless, the qualitative picture of the physics we observe, up to the intermediate length and timescales considered in this paper, is not likely to change. Indeed, it is reasonable to expect that vison-vison interactions only alter the entropic term of the effective free energy quantitatively but not qualitatively, whereas spinons, in the low-temperature low-density regime, are separated by length scales much longer than any reasonable characteristic interaction range, and their behavior should remain unchanged up to reasonably long timescales.

We also briefly discussed the nonequilibrium behavior of the system when the temperature is varied in time at some finite rate. We expect a similar hysteretic behavior as in 2D~\cite{MainCorr}. When the system is cooled, the spinon density $\rho_s$ falls off equilibrium and forms a plateau, before it eventually becomes kinematically locked to $\rho_s \sim T^{3/5}$ when the vison-depleted regions become large enough to overlap (i.e., their size is approximately equal to the characteristic spinon separation length). 

Our results demonstrate how the rich phenomenology at the interface between topological order and Anderson localization in disorder-free systems arises also in higher dimensional systems. Whereas the localization properties are different, the notable effects on the relaxation and transport properties underpinned by the formation of vison depletion regions around the spinons survive. This behavior is driven by the mutual semionic statistics and can thus be taken as a signature of 3D QSL behavior at finite temperature. The hysteretic behavior discussed in Sec.~\ref{sec:out-of-equilibrium} could be probed, for instance, with techniques that access spinon density, which is expected, for example, to directly affect the magnetic susceptibility~\cite{MainCorr}. The presence of vison depletion regions bound to the spinons is also likely to have distinctive repercussions on transport properties where spinons or visons contribute (e.g., thermal transport). An interesting future direction could be to model more extensively the dynamical interplay of spinons and visons, including spinon annihilation events when vison depletion regions overlap, and possibly simulating the out-of-equilibrium behavior directly. Another direction could be to extend our analysis to other classes of Hamiltonians such as $\mathbb{Z}_n$ gauge theories where the effective fluxes break time-reversal symmetry, and to general 3D string-net models~\cite{Stringnet, 3DToric}. 
As we await the discovery of candidate materials that realize $\mathbb{Z}_2$ QSL phases in 3D, our results may be relevant to other contexts, including frustrated magnetic pyrochlore oxides and resonant valence bond systems. Some of these systems exhibit further gapless excitations, and it would be interesting to see how the behavior discussed in this paper is affected by their presence. 
Moreover, the possibility of realizing $\mathbb{Z}_2$ spin liquid Hamiltonians in our temperature regime with quantum annealers~\cite{QSLConstruction,QuantumAnnealExp1} and quantum simulators~\cite{Simulator1,Simulator2} (albeit typically limited to 2D) could provide a suitable arena where the physics discussed here could be tested and explored further.
%
%
\begin{acknowledgements}
We are very grateful to O. Hart for the generous guidance in the early stages of this project and for several useful discussions thereafter. 
This paper was supported in part by the Engineering and Physical Science Research Council (EPSRC) Grants No.~EP/P034616/1, No.~EP/T028580/1, and No.~EP/V062654/1 (CC). GDT acknowledges support from the EPiQS Program of the Gordon and Betty Moore Foundation.
MK developed and performed the calculations and numerical simulations. 
\end{acknowledgements}
%
%
\appendix 
\section{\label{app:statistics}
Further properties of localized and extended states
        }
In this appendix, we present details of the study of the DOS and the scaling properties of the IPR for the localized and extended states, and we further discuss the behavior of the probability distributions and level statistics in regions around the mobility edge to complement our discussion in Sec.~\ref{sec:mobility}. 

\begin{figure}[ht!]
    \centering
    \includegraphics[width=0.98\columnwidth]{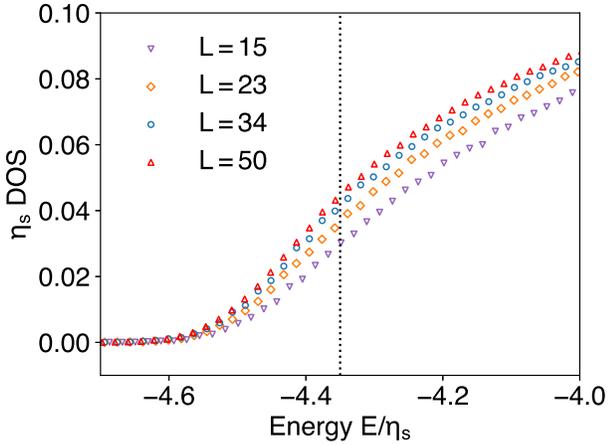}
    \caption{Density of states (DOS) for various system sizes. The mobility edge, being close to the band edge, is on the region where the DOS profile is not flat. We note the generally higher DOS for larger system sizes closer to the band tail, which is compensated by the relative depression near the band center. The dashed vertical line indicates the mobility-edge and is a guide for eye.}
    \label{fig:dos}
\end{figure}

In the main text, we determined the mobility edge to be around $E/\eta_s \sim -4.35$, which is close to the band edge, as shown in Fig.~\ref{fig:dos}.
This leads to a suppression of the DOS in the localized region of the spectrum and a relatively quickly varying DOS near the mobility edge. The smaller DOS also leads to larger Mott temperature, making finite-temperature dephasing effects relevant only at long timescales, as discussed in the main text.

\begin{figure}[ht!]
\centering
\includegraphics[width=0.45\columnwidth]{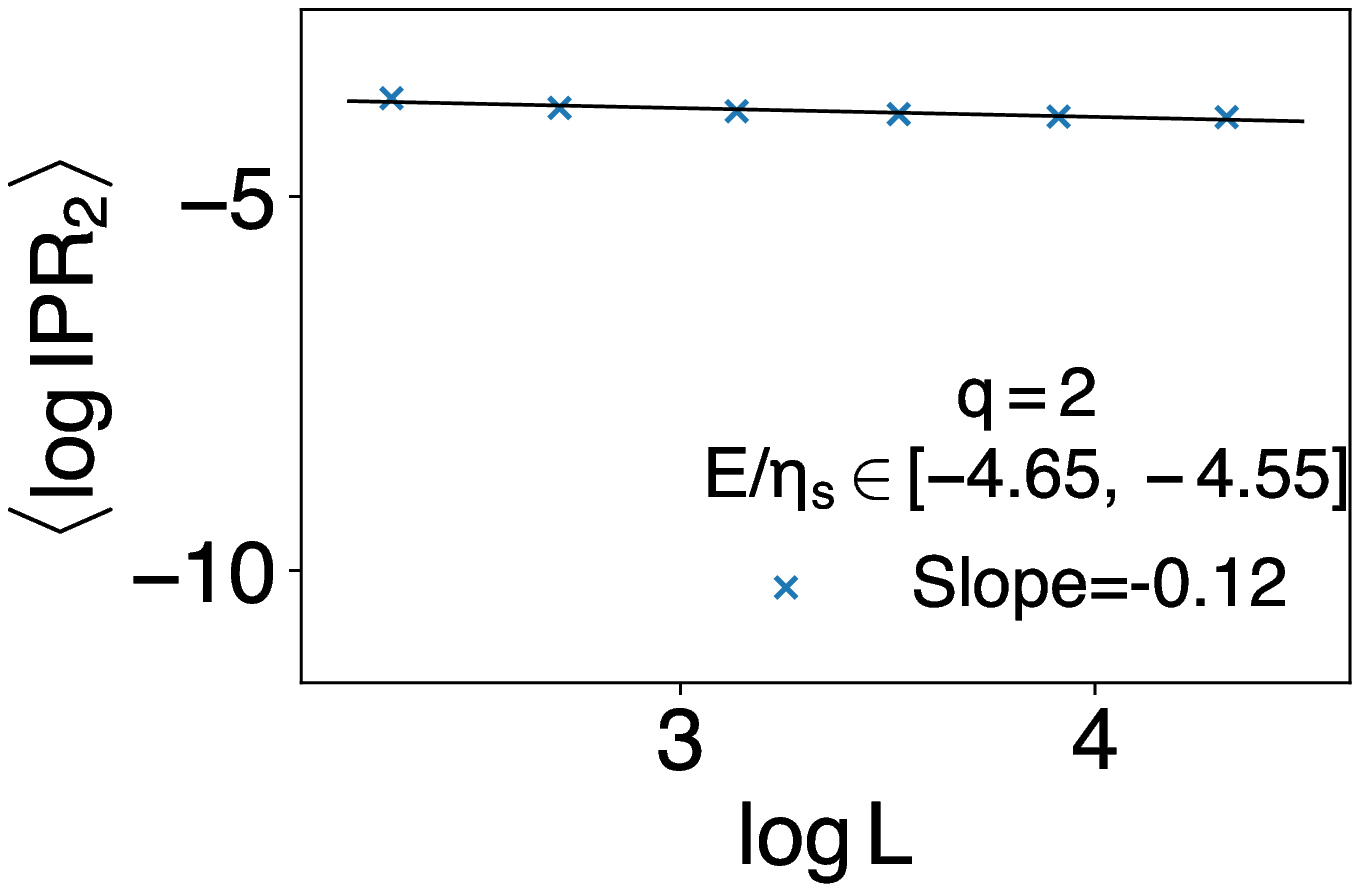}
\includegraphics[width=0.45\columnwidth]{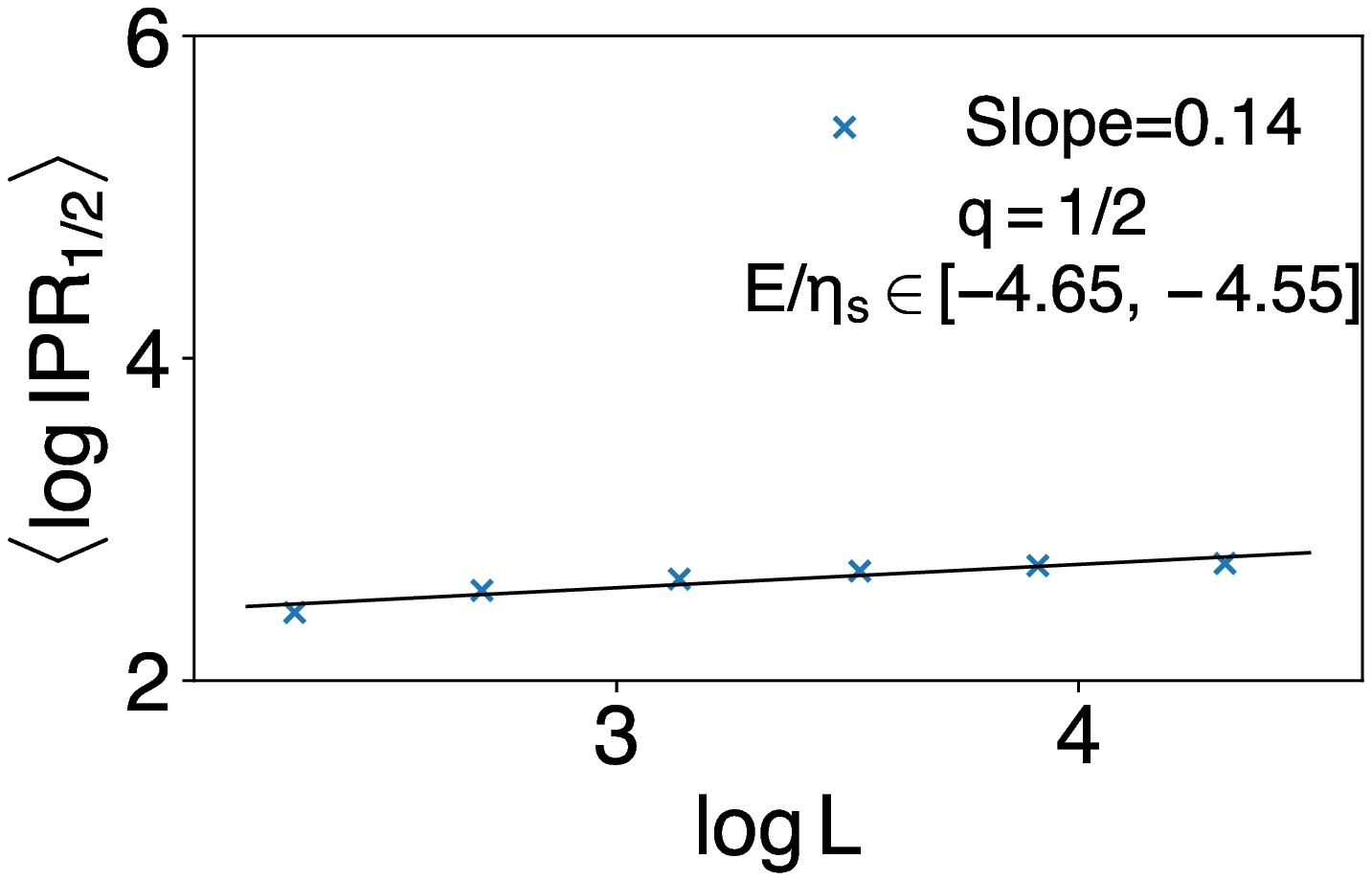}
\\
\includegraphics[width=0.45\columnwidth]{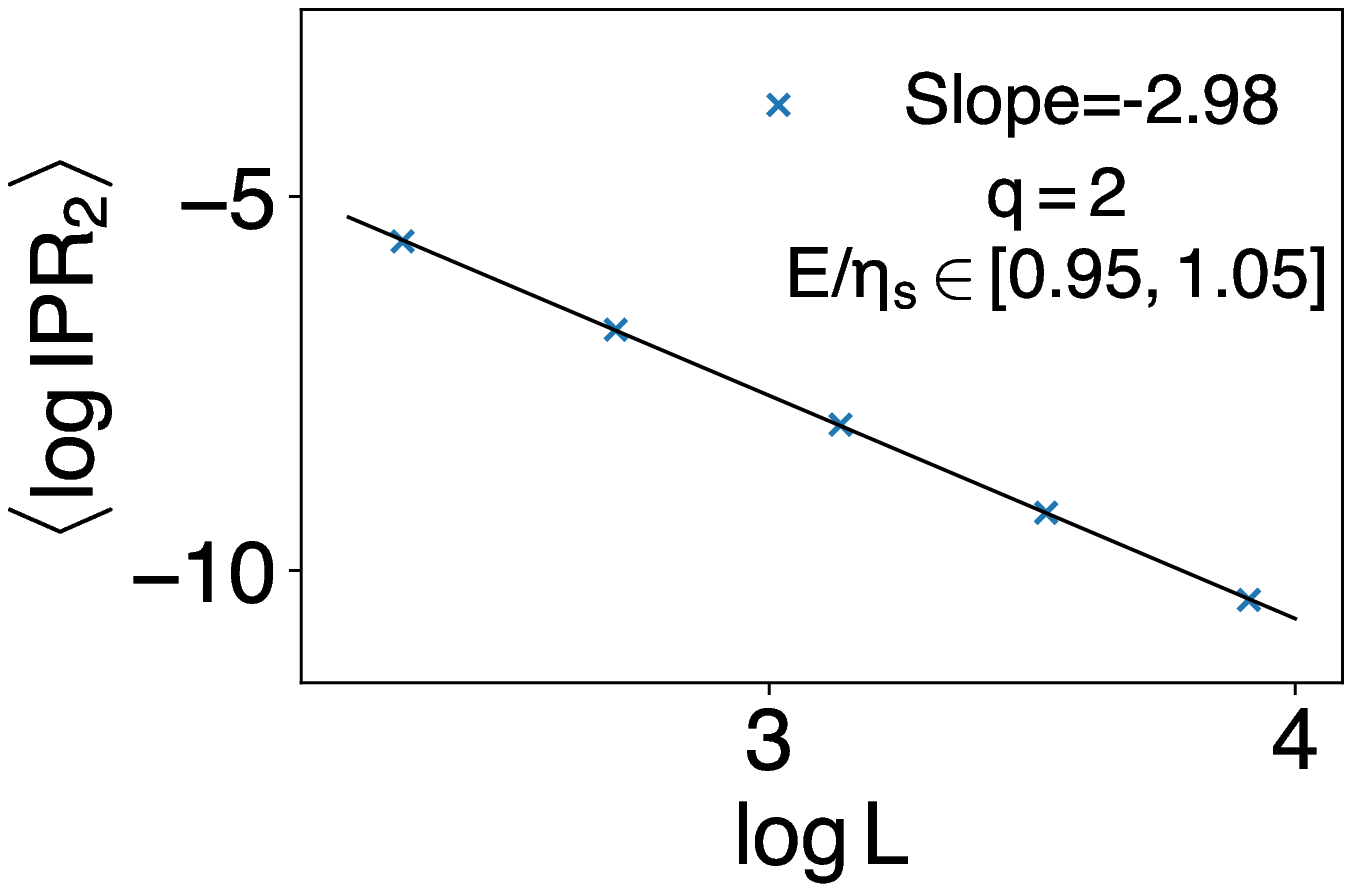}
\includegraphics[width=0.45\columnwidth]{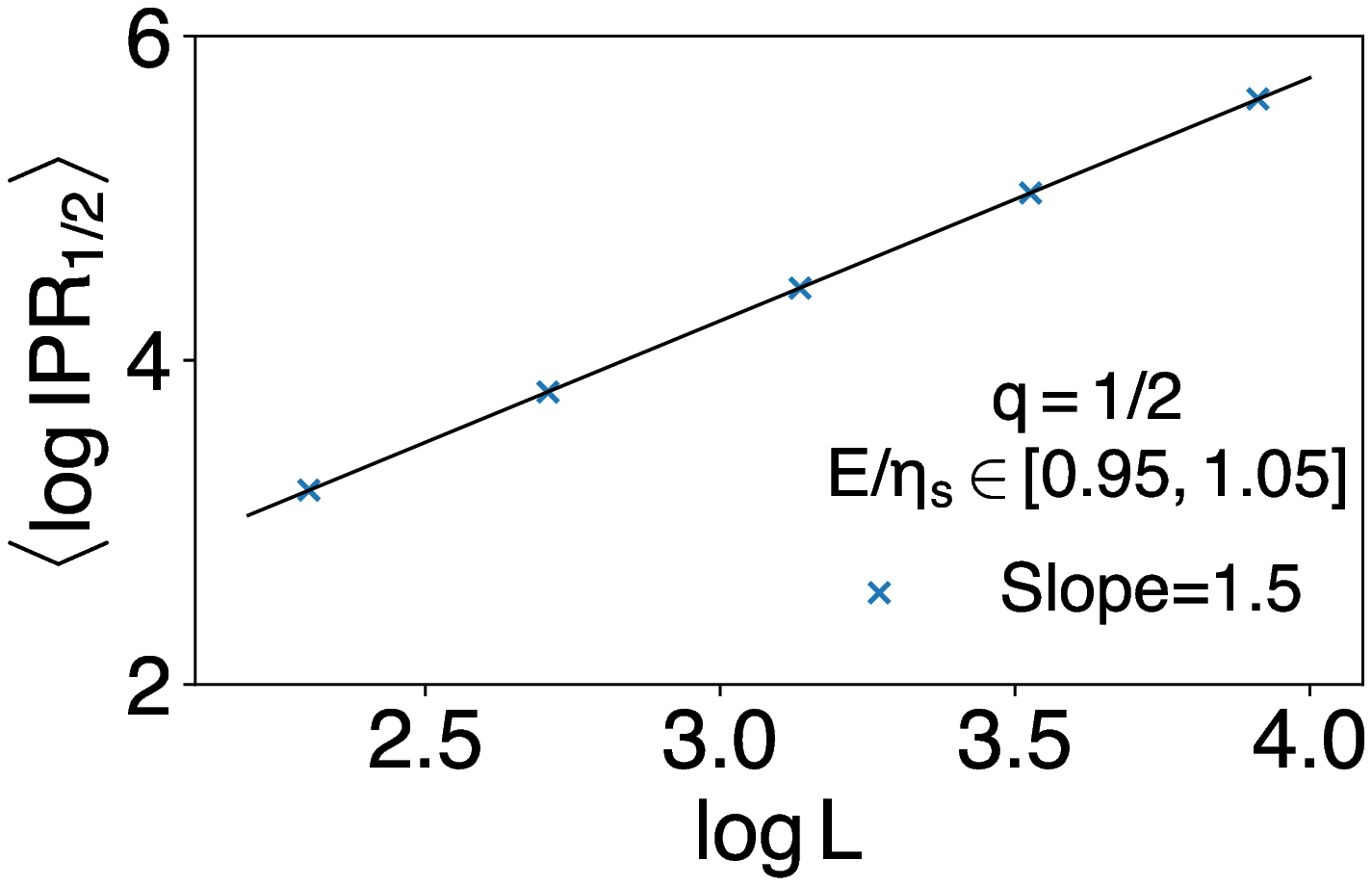}
\caption{\label{fig:IPRscaling}
Scaling with system size of the IPR$_q$ for localized (top panels) and delocalized (bottom panels) eigenstates, for $q=2$ (left panels) and $q=\frac{1}{2}$ (right panels). For localized states, $\ln {\rm IPR}_q$ are sampled from states in the energy window $E/\eta_s \in [-4.65, -4.55]$ in units of the spinon hopping. The extended states are sampled in the window $E/\eta_s \in [0.95, 1.05]$.
}
\end{figure}

In Fig.~\ref{fig:IPRscaling}, we show directly the scaling of IPR$_q$, contrasting states in the localized vs delocalized portions of the energy spectrum.
The localized states are consistent with IPR$_q$ taking a constant value (the fitted slope being close to zero), whereas the delocalized states are in very good agreement with the expected scaling, $\ln \langle {\rm IPR}_q \rangle \sim 3(1-q)\ln L$. 

\begin{figure}[ht!]
\centering
\includegraphics[width=0.45\columnwidth]{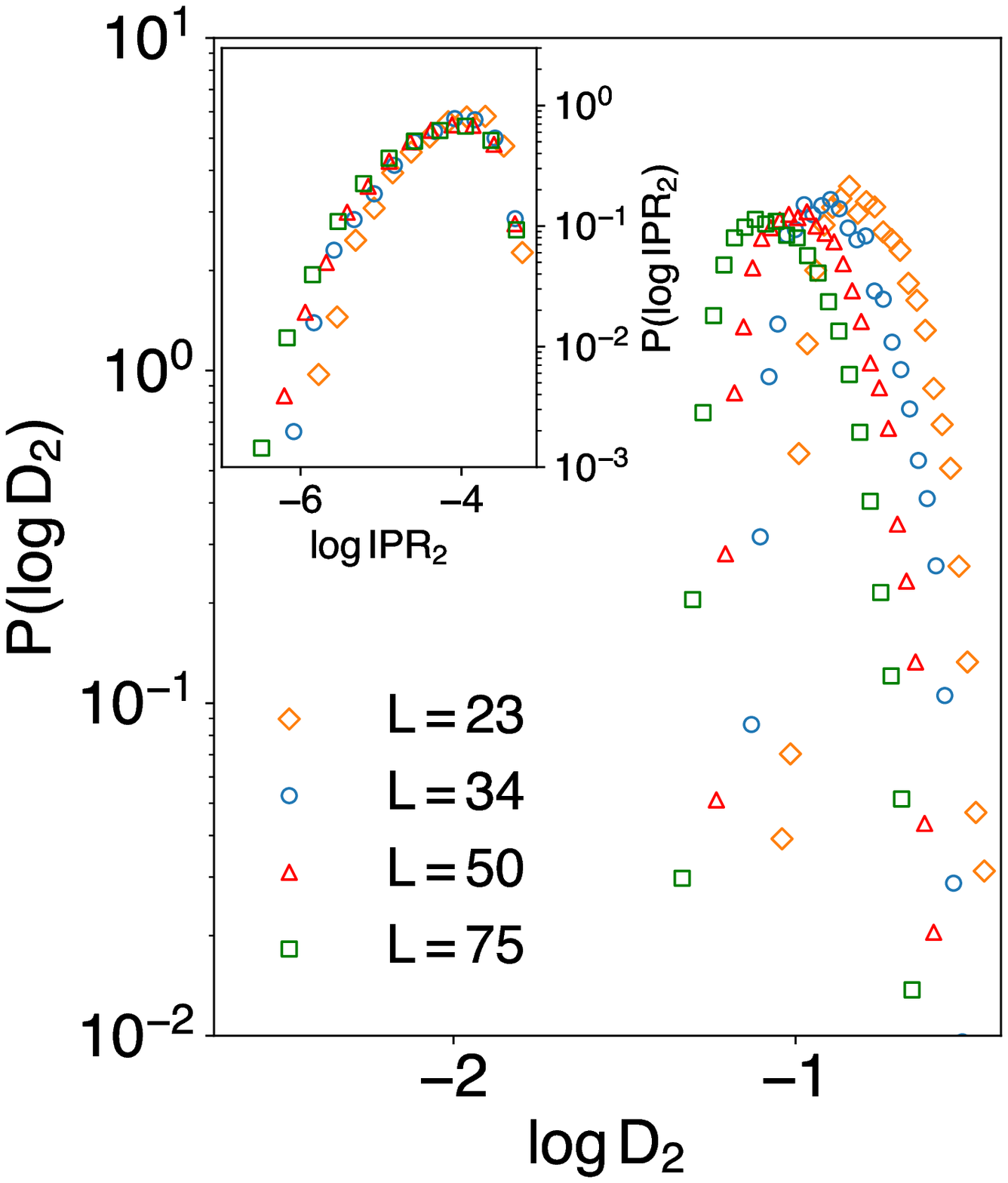}
\includegraphics[width=0.45\columnwidth]{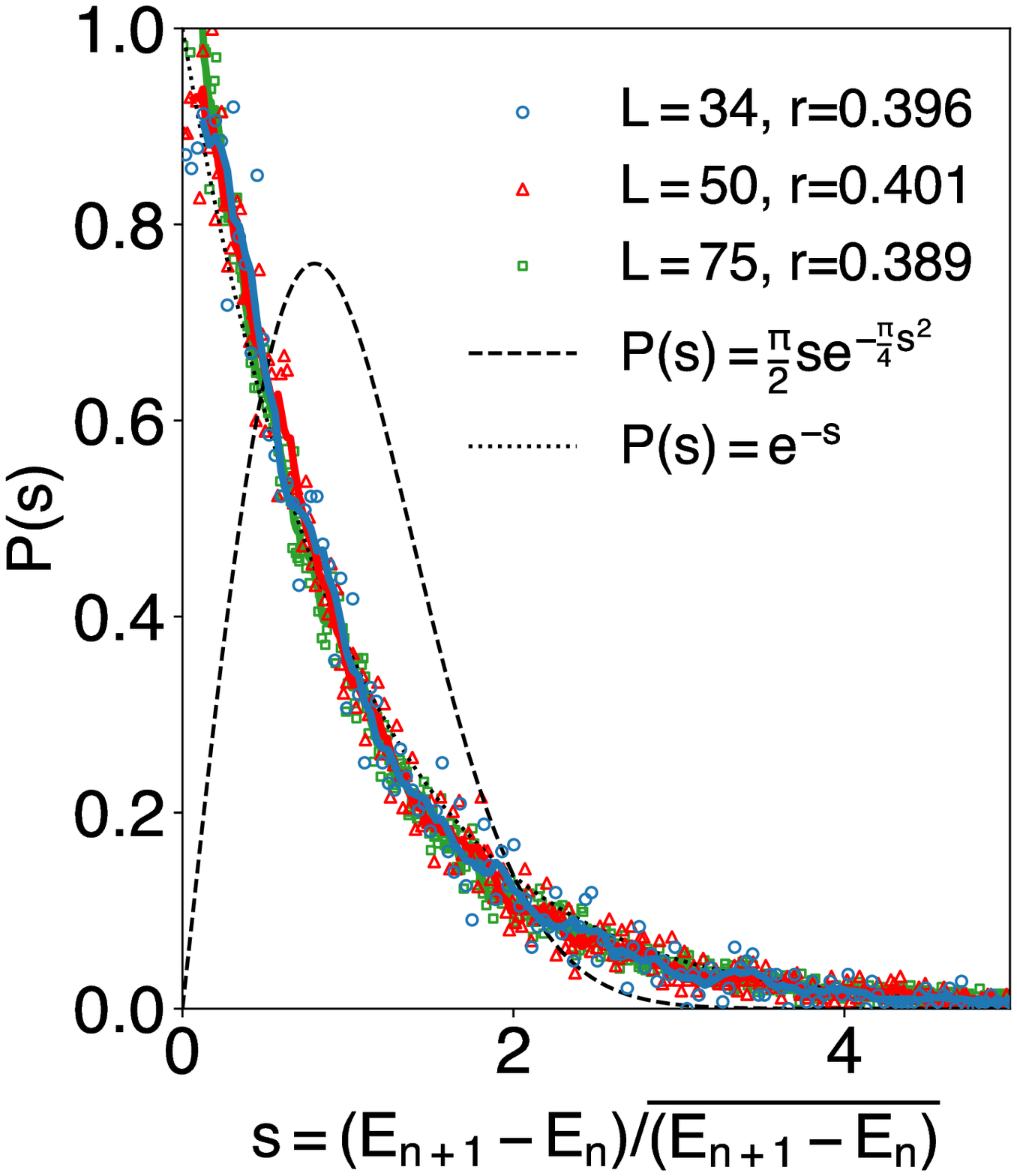}
\\
\includegraphics[width=0.45\columnwidth]{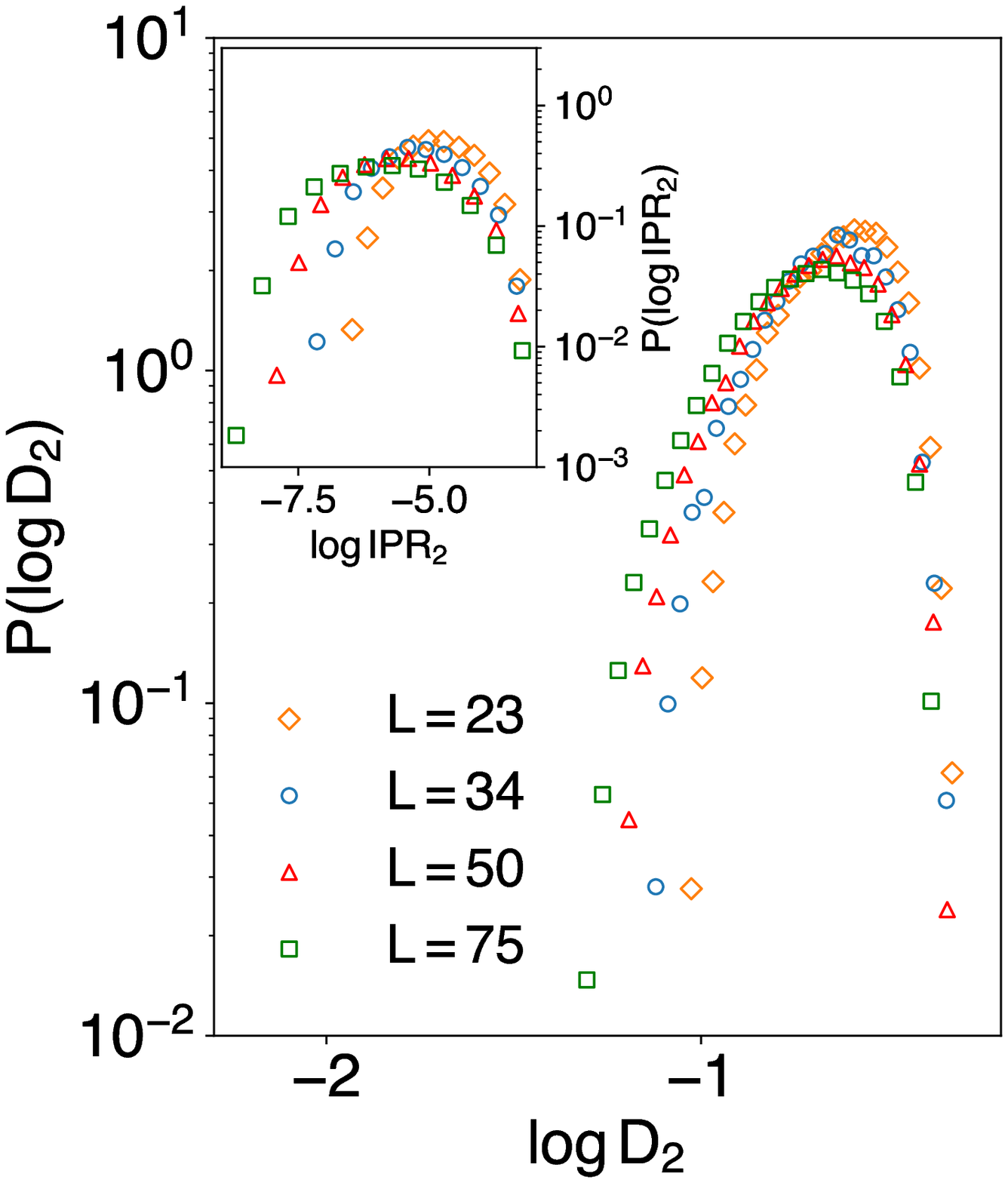}
\includegraphics[width=0.45\columnwidth]{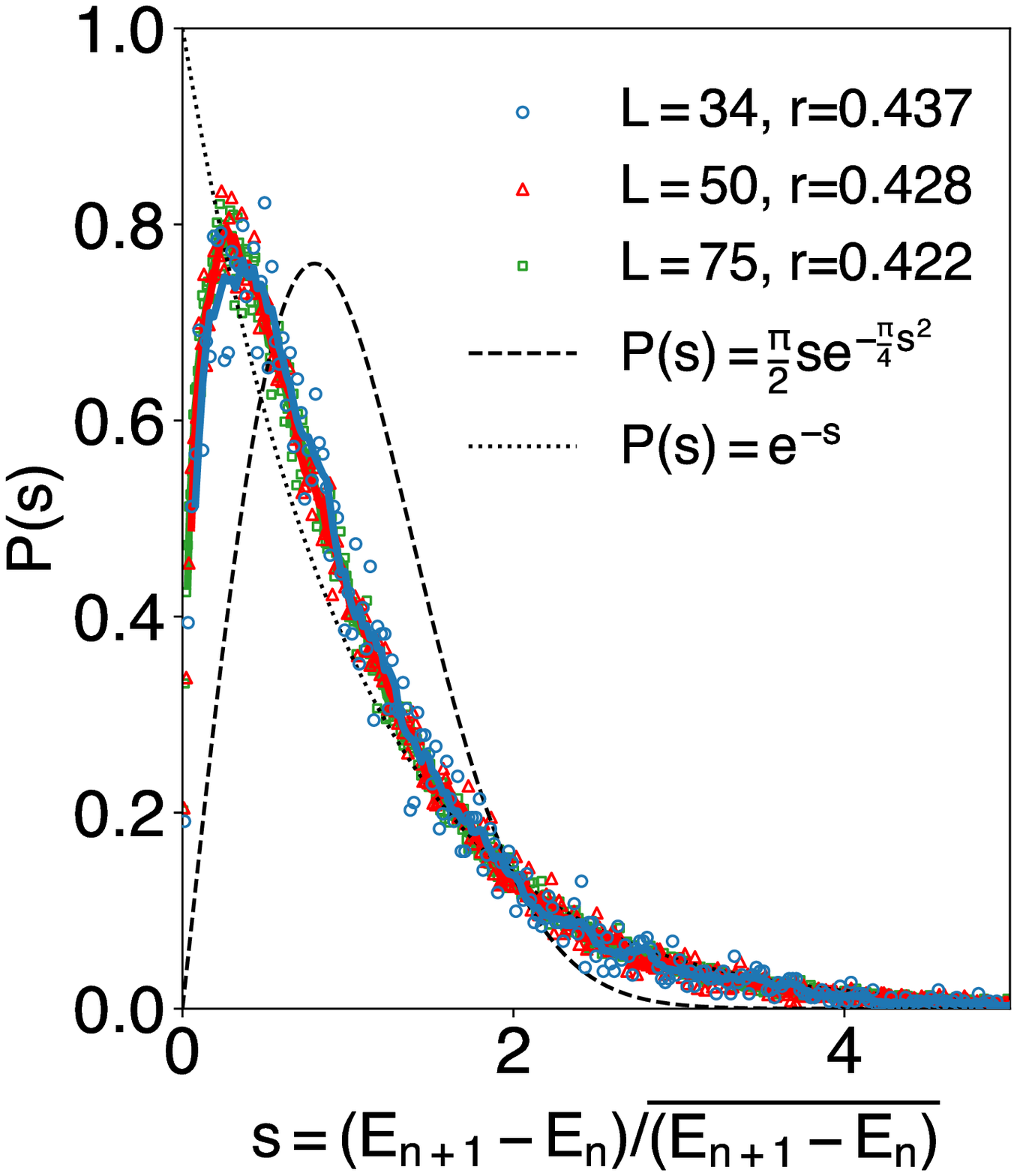}
\\
\includegraphics[width=0.45\columnwidth]{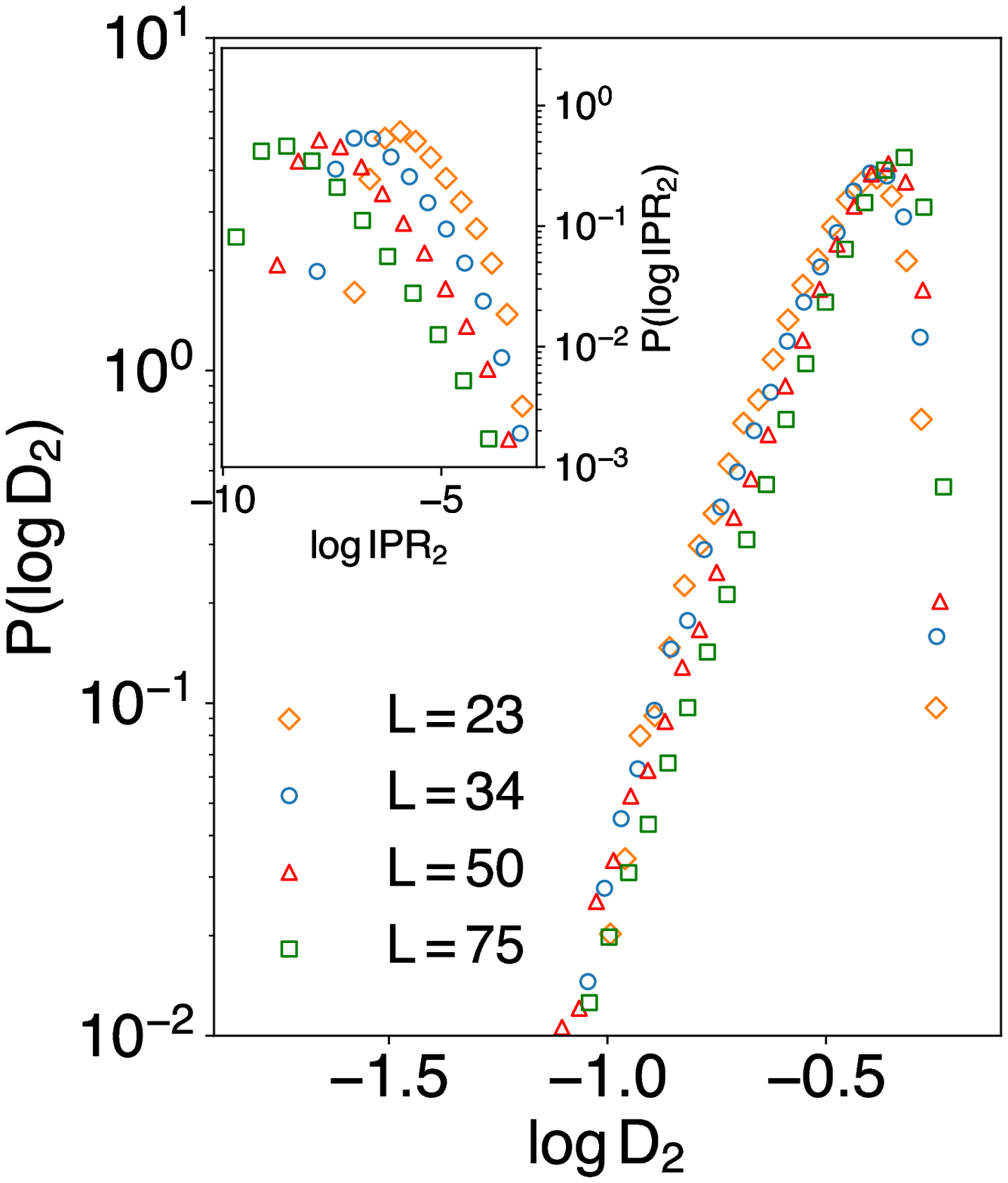}
\includegraphics[width=0.45\columnwidth]{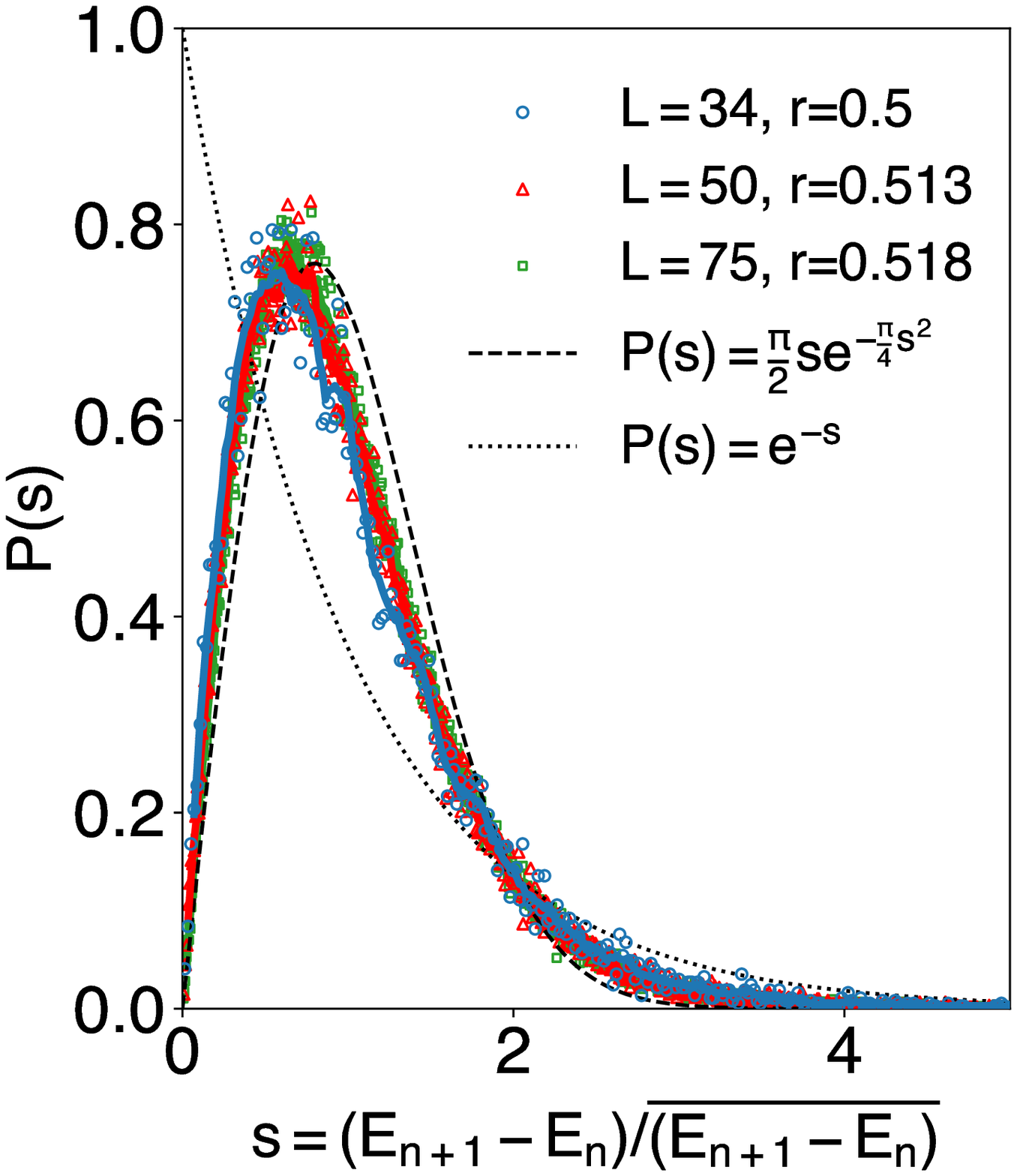}
\caption{\label{fig:D2IPRscaleinter}
Probability distributions for the multifractal coefficient $D_2$ (left panels) and the IPR$_2$ (insets), alongside the level statistics (right panels), for a fixed energy window of width $0.1 \eta_s$ placed in different regions of the spectrum, spanning localized and extended states. From top to bottom, the values of the energy in the middle of each window are $E/\eta_s = -4.5$, $-4.4$, $-4.3$, respectively. We note a semi-Poissonian distribution appearing just before the mobility edge is reached, due to the fact that the probability distributions of the IPR do not exactly align with each other at energies lower but close to the mobility edge (e.g., near $E/\eta_s \sim -4.4$).}
\end{figure}

Near the mobility edge, the level statistics and the probability distributions of IPR$_2$ and $D_2$ gradually change from localized to extended behavior as we sample higher energy states. This is illustrated in Fig.~\ref{fig:D2IPRscaleinter} for a range energies from localized and extended. 

As the energy increases from the lowest end of the spectrum, the level statistics transitions from a Poissonian distribution to a Wigner surmise, via a semi-Poissonian regime~\cite{semipoisson, Evangelou2003}. 
The transition starts from around $E/\eta_s \sim -4.5$, which indicates the existence of level repulsion even in regions slightly below the mobility edge. The emergence of the semi-Poissonian is due to the fact that the states are not fully localized and it is the typical behavior around the Anderson transition~\cite{Shklovskii_93}.

Correspondingly, around this energy, the tail ends of the probability distributions of the IPR for different system sizes start to deviate from each other. The behavior of the multifractal coefficient distribution is like IPR$_2$, given that the shape of the distribution curves starts to transition at similar energies. 
These results are consistent with the $r$ statistics (see also Fig.~\ref{fig:rstat}). 

\begin{figure}[ht!]
\centering
\includegraphics[width=0.45\columnwidth]{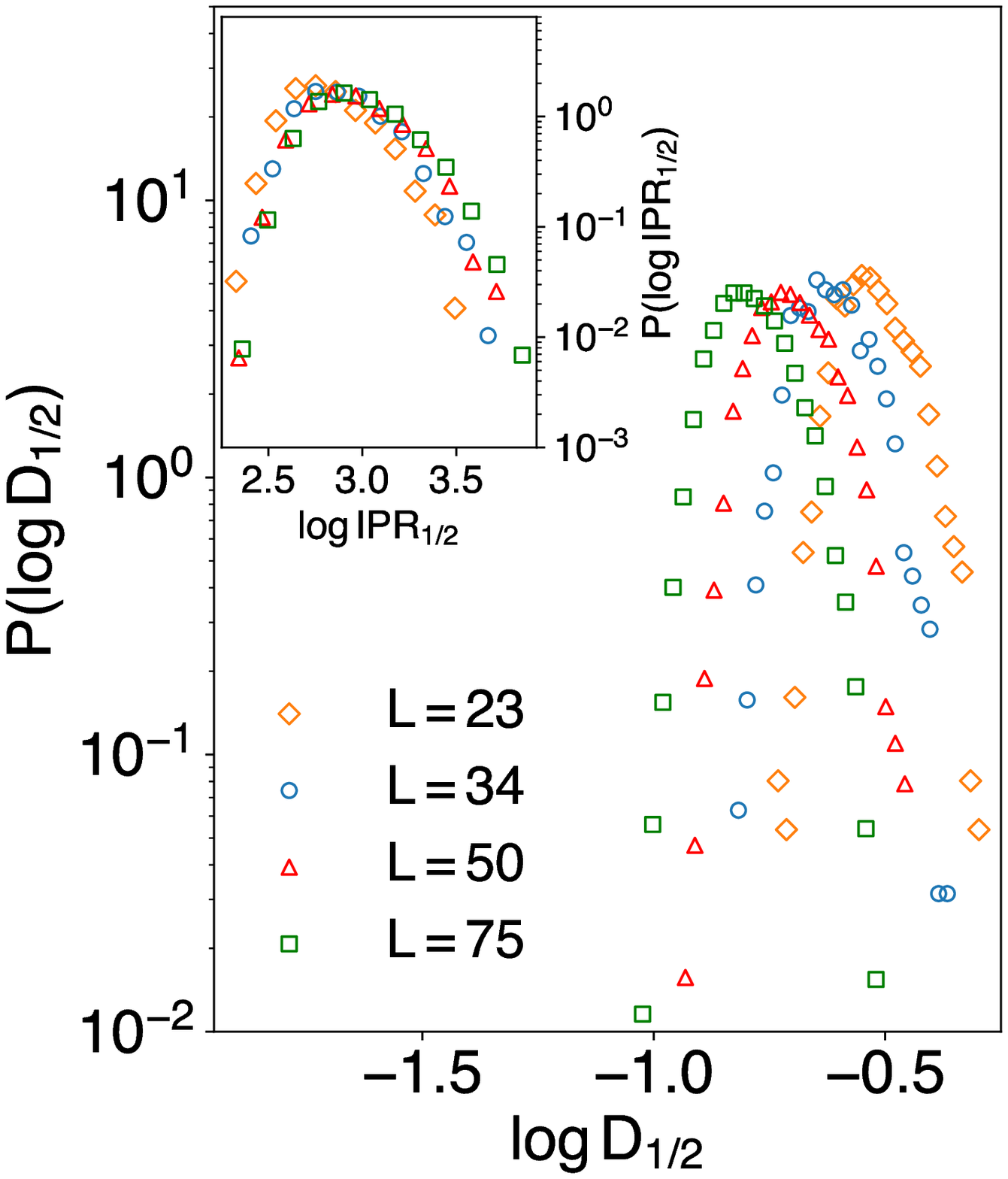}
\includegraphics[width=0.45\columnwidth]{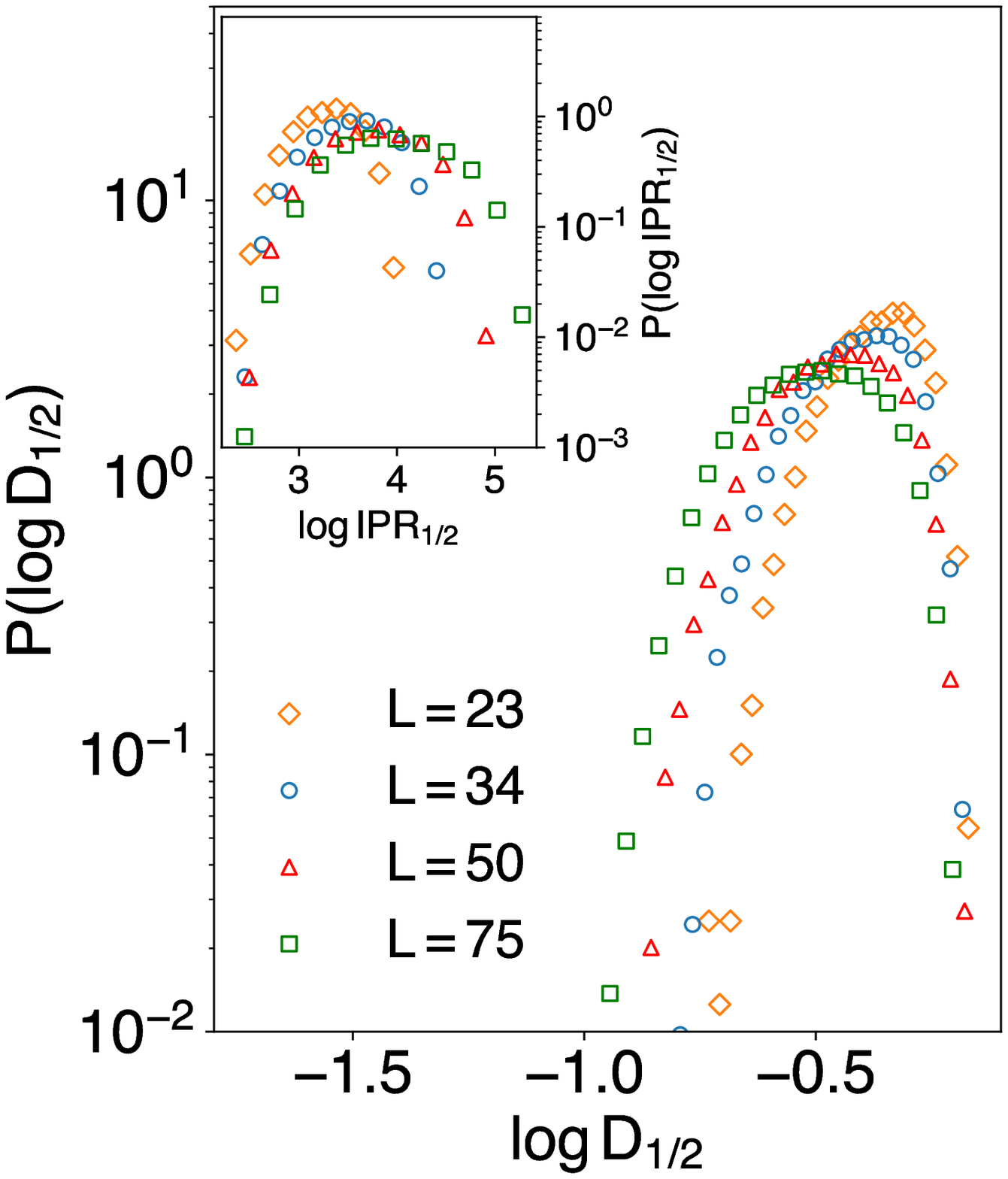}
\\
\includegraphics[width=0.45\columnwidth]{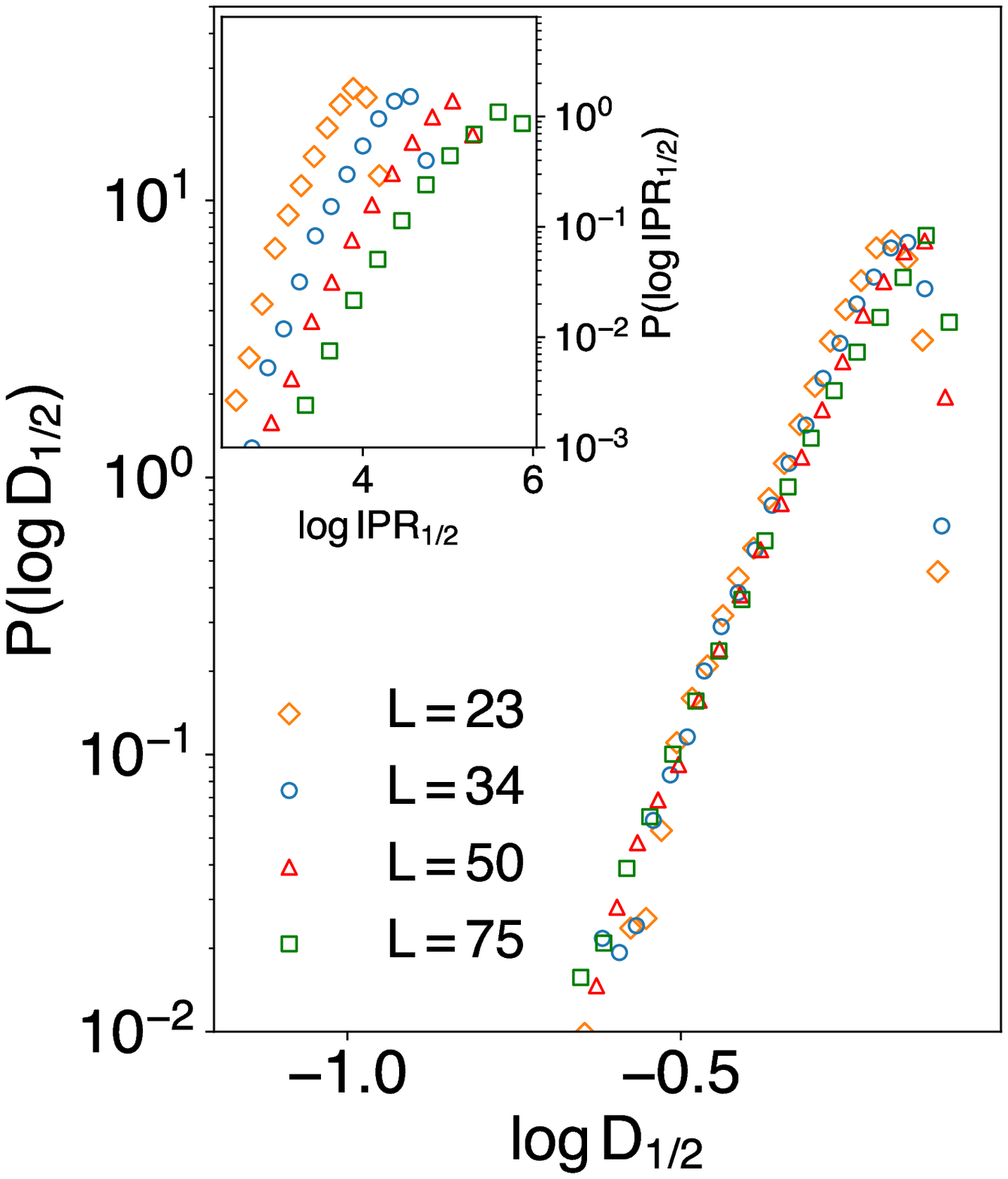}
\includegraphics[width=0.45\columnwidth]{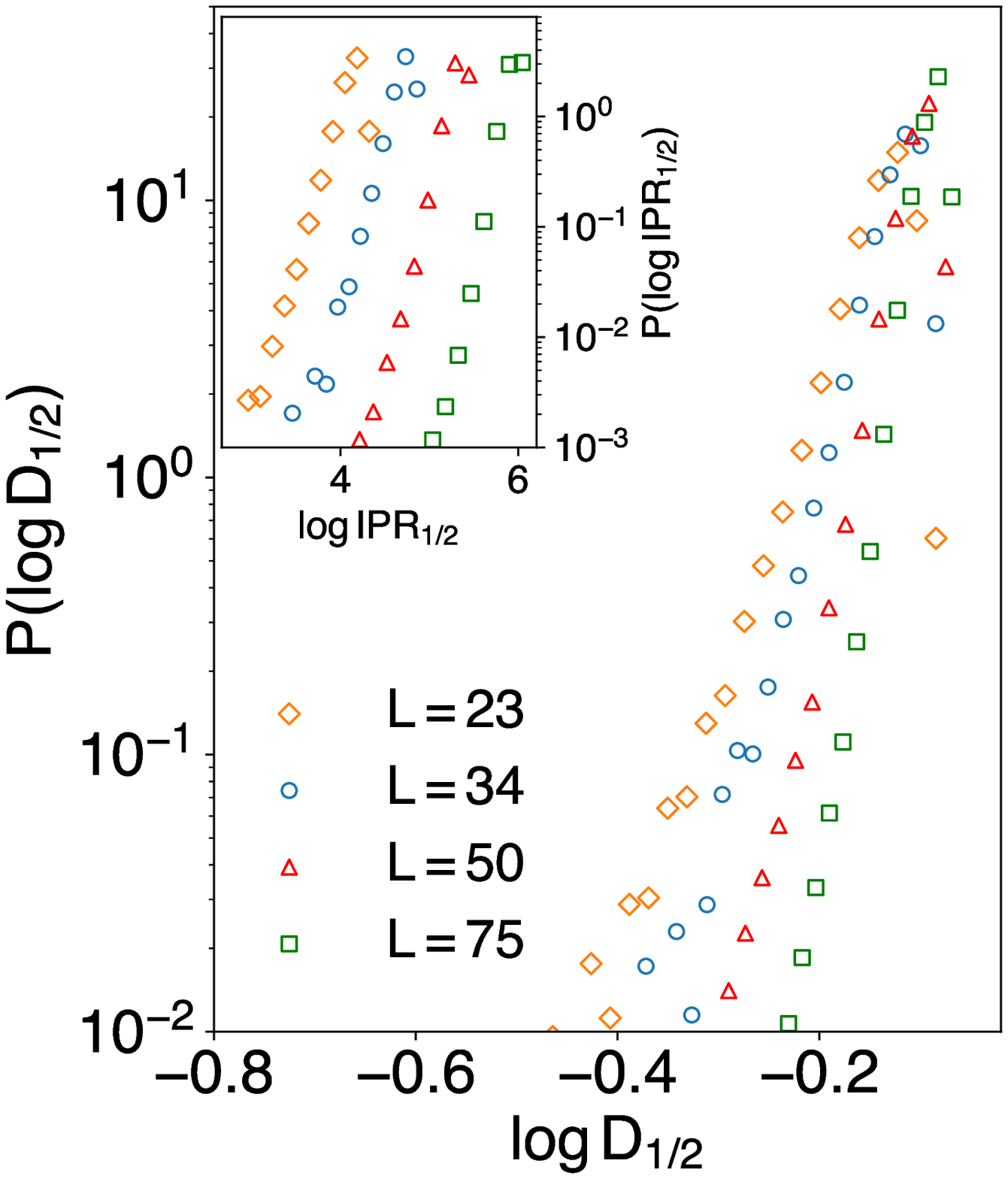}
\caption{\label{fig:IPRscaleinter2} Probability distributions for the multifractal coefficient and the IPR for $q=\frac{1}{2}$, for a range of energies between $E/\eta_s = -4.55$ to $-4.25$, with energy spacing $0.1 \eta_s$ and energy window width of $0.1 \eta_s$. The energies increase from left to right and from top to bottom. We see an overall similar pattern with $D_2$ and IPR$_2$. 
}
\end{figure}

Figure~\ref{fig:IPRscaleinter2} shows the probability distribution results for $D_{\frac{1}{2}}$ and IPR$_{\frac{1}{2}}$, which exhibit the same behavior as $q=2$ above.
%
%

\section{\label{app:Bethe}
Bethe lattice calculation
        }
Following the Bethe lattice calculation in Ref.~\onlinecite{MainProp}, we derive similar results for our model defined on a 3D cubic lattice. 
We start from the generating functions for closed and open walks on a Bethe lattice of coordination number $z=6$ (or branching ratio $z-1$), which are given as
\begin{eqnarray}
  T^{(z)}(x) &=& \frac{2(z-1)}{z-2+z \sqrt{1-4(z-1)x^2}},
\\
  S^{(z)}(x) &=& \frac{1-\sqrt{1-4(z-1) x^{2}}}{2(z-1) x}
  \, . 
\end{eqnarray}
We can also directly use the analytical form for the $2k$th moment of the displacement:
\begin{eqnarray}
\left\langle\mathbf{r}^{2 k}(t)\right\rangle &=& \oiint \frac{d w_{1}}{2 \pi i} \frac{d w_{2}}{2 \pi i} \frac{\exp[i h t\left(w_{1}-w_{2}\right)]}{w_{1} w_{2}} T\left(\frac{1}{w_{1}}\right) T\left(\frac{1}{w_{2}}\right) \nonumber \\
&\times& \mathcal{R}_{2 k}\left[S\left(\frac{1}{w_{1}}\right) S\left(\frac{1}{w_{2}}\right)\right]
\, . 
\label{Bethe_moment}
\end{eqnarray}

The nonreversing constraints, however, need to be modified. Unlike the 2D $z=6$ case, the constraints now have three variables, as 3D lattice walks require three independent vectors. We introduce the generating variables $x, \delta, \epsilon$, and $\zeta$ which count the length of the walk and the number of steps taken in the lattice directions $\mathbf{e}_{1}, \mathbf{e}_{2}$, and $\mathbf{e}_{3}$, respectively. At each lattice point, there are six possible walks: $\mathbf{e}_{1}, \mathbf{e}_{2}, \mathbf{e}_{3}$ and their reverses, each noted by $\delta, \epsilon, \zeta, \delta^{-1}, \epsilon^{-1}$, and $\zeta^{-1}$, respectively, in generating variables. In addition to the first step, however, the immediate reversing of the previous step is forbidden. We can denote this in a matrix, given as 
\begin{equation}
N=x\left(\begin{array}{cccccc}
\delta & \epsilon & \zeta & \zeta^{-1} & \epsilon^{-1} & 0 \\
\delta & \epsilon & \zeta & \zeta^{-1} & 0 & \delta^{-1} \\
\delta & \epsilon & \zeta & 0 & \epsilon^{-1} & \delta^{-1} \\
\delta & \epsilon & 0 & \zeta^{-1} & \epsilon^{-1} & \delta^{-1} \\
\delta & 0 & \zeta & \zeta^{-1} & \epsilon^{-1} & \delta^{-1} \\
0 & \epsilon & \zeta & \zeta^{-1} & \epsilon^{-1} & \delta^{-1}
\end{array}\right),
\end{equation}
with the initial condition $N_{0}=x \operatorname{diag}\left(\delta, \epsilon, \zeta, \zeta^{-1}, \epsilon^{-1}, \delta^{-1}\right)$. We note that, for each step, the length of the path is advanced by one, and the matrix multiplication ensures that all possible combinations of steps are accounted for. The zero entries represent the nonreversing constraint, while $N_{0}$ is unconstrained.

The matrix elements $N_0 N^l$ therefore give the paths of length $l$ that are consistent with the nonreversing constraint. The generation function $\mathcal{N}(x ; \delta, \epsilon, \zeta)$ for nonreversing paths can then be defined as the sum over all matrix elements and all possible path lengths $l$, including also the trivial walk of zero length. This can be written as

\begin{equation}
\begin{split}
\mathcal{N}(x ; \delta, \epsilon, \zeta) &=1+\sum_{i, j} \sum_{\ell=1}^{\infty}\left[N_{0} N^{\ell-1}\right]_{i j} \\
&=1+\sum_{i, j}\left[N_{0}\left(\mathbbm{1}_{z}-N\right)^{-1}\right]_{i j} \\
&=\frac{1-x^{2}}{1+5 x^{2}-x\left(\delta+\delta^{-1}+\epsilon+\epsilon^{-1}+\zeta+\zeta^{-1}\right)}
\, . 
\end{split}
\end{equation}

The generator for the $2k$th moment of the displacement therefore is
\begin{equation}
\begin{split}
\mathcal{R}_{2 k}(x) & \equiv \sum_{s \in \mathcal{L}_{z}}\left(s_{x}^{2}+s_{y}^{2}+s_{z}^{2}\right)^{k} C_{s}(x) \\
&=\left.\left\{\left[\left(\delta \partial_{\delta}\right)^{2}+\left(\epsilon \partial_{\epsilon}\right)^{2}+(\zeta \partial_{\zeta})^{2}\right]^{k} \mathcal{N}(x ; \delta, \epsilon, \zeta)\right\}\right|_{\delta=\epsilon=\zeta=1}
\, . 
\end{split}
\end{equation}
For $k=1$, the generator takes the simpler form:
\begin{equation}
\mathcal{R}_{2}(x) = \frac{6 x(1+x)}{(1-5 x)^{2}(1-x)}
\, . 
\end{equation}

When we substitute the relevant generators into Eq.~\eqref{Bethe_moment}, we notice that there is a line of poles due to $S\left(u+i 0^{+}\right) S\left(u-i 0^{+}\right)=(z-1)^{-1}$ for $u \in \mathbb{R}$ and $|u|<2 \sqrt{z-1}$. 
The pole signature and positions are the same as in 2D; hence, we can directly use
\begin{equation}
    2 D_{z}^{\mathrm{diff}}=\frac{1}{2 \pi} \mathcal{F}(z) \lim _{w \rightarrow(z-1)^{-1}}\left[1-w(z-1)\right]^{2} \mathcal{R}_{2}^{(z)}(w)
\, , 
\end{equation}
where 
\begin{equation}
\begin{split}
\mathcal{F}(z) & \equiv(z-1) \int_{-2 \sqrt{z-1}}^{2 \sqrt{z-1}} ~du \frac{4(z-1)-u^{2}}{z^{2}-u^{2}} \\
&=(z-1)\left[4 \sqrt{z-1}-z\left(\frac{z-2}{z}\right)^{2} \ln \left(\frac{z+2 \sqrt{z-1}}{z-2 \sqrt{z-1}}\right)\right],
\end{split}
\end{equation}
which is the quantity we obtain after integrating over the relevant residue line.

The final asymptotic diffusion coefficient is given by
\begin{equation}
D_{6}^{\mathrm{diff}} 
= \frac{3}{\pi}\left[3 \sqrt{5}-2 
\ln \left(\frac{3+\sqrt{5}}{3-\sqrt{5}}\right)\right] 
\simeq 2.72968
\, , 
\end{equation}
which is coincidentally identical to the case of a triangular lattice in 2D. 
%
%

\section{\label{app:strip}
Low temperature finite-size instabilities of the vison depletion region
        }
\begin{figure}[ht!]
    \centering
    \subfloat{
		\includegraphics[width=0.47\linewidth]{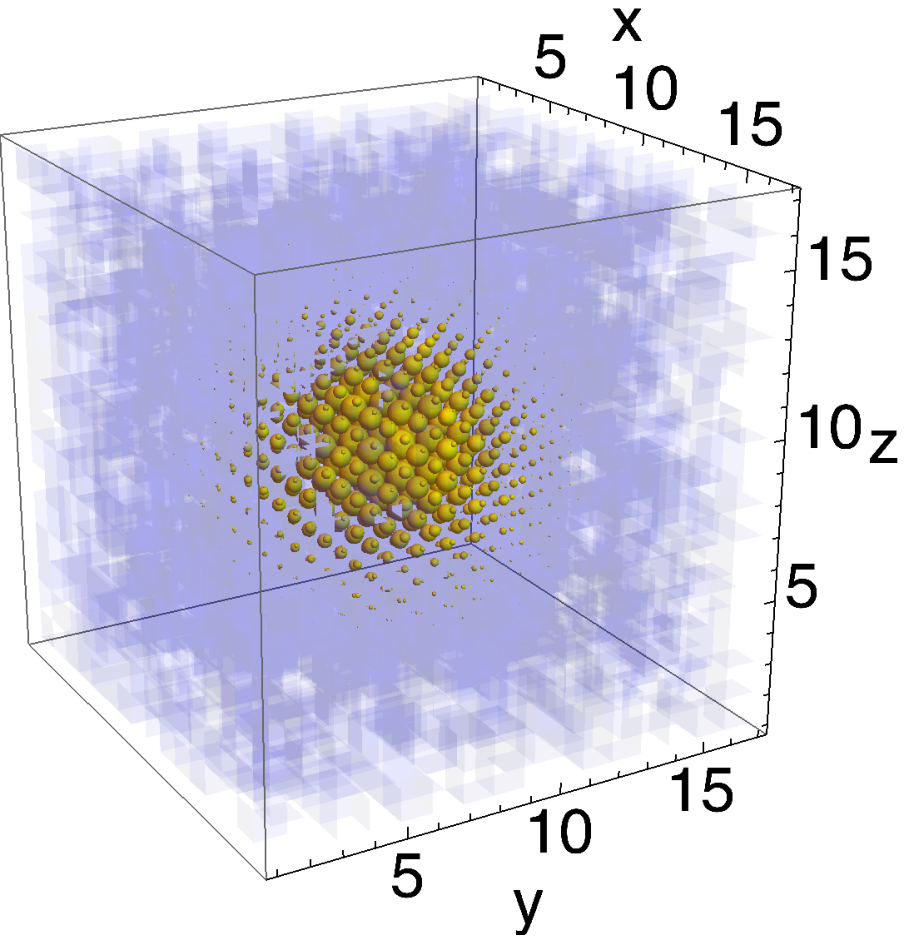}
    }
    \subfloat{
    \includegraphics[width=0.47\linewidth]{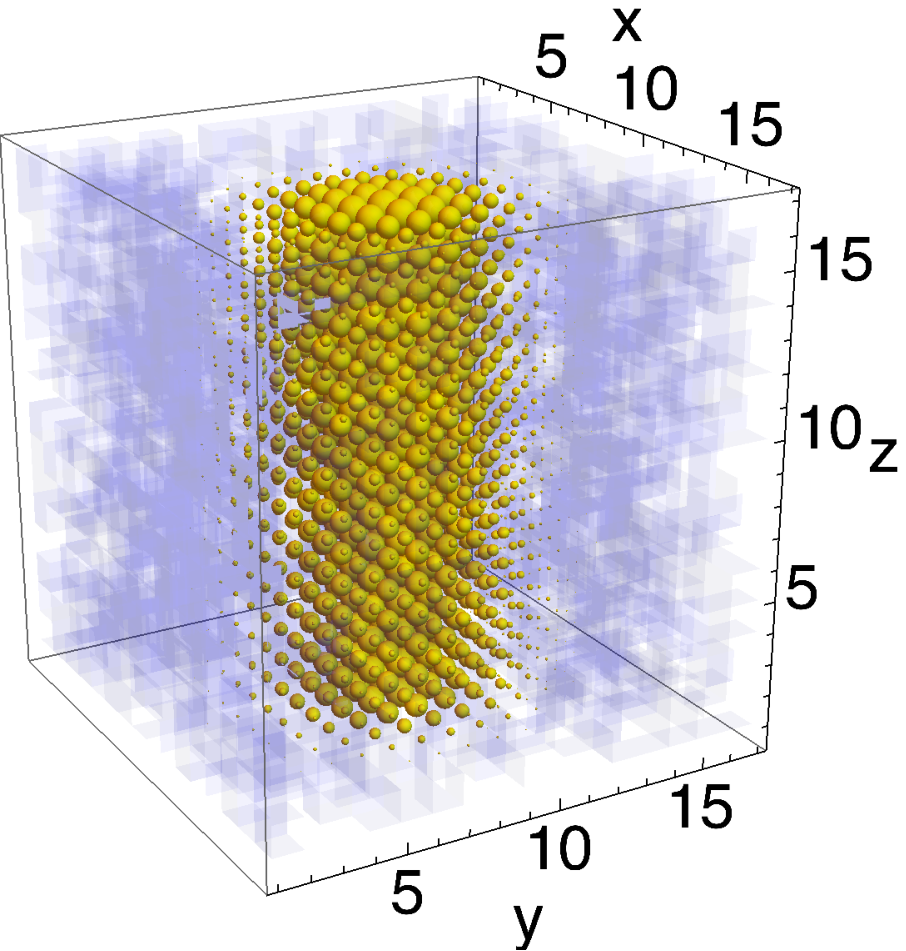}
    }
    \vskip\baselineskip
    \subfloat{
    \includegraphics[width=0.47\linewidth]{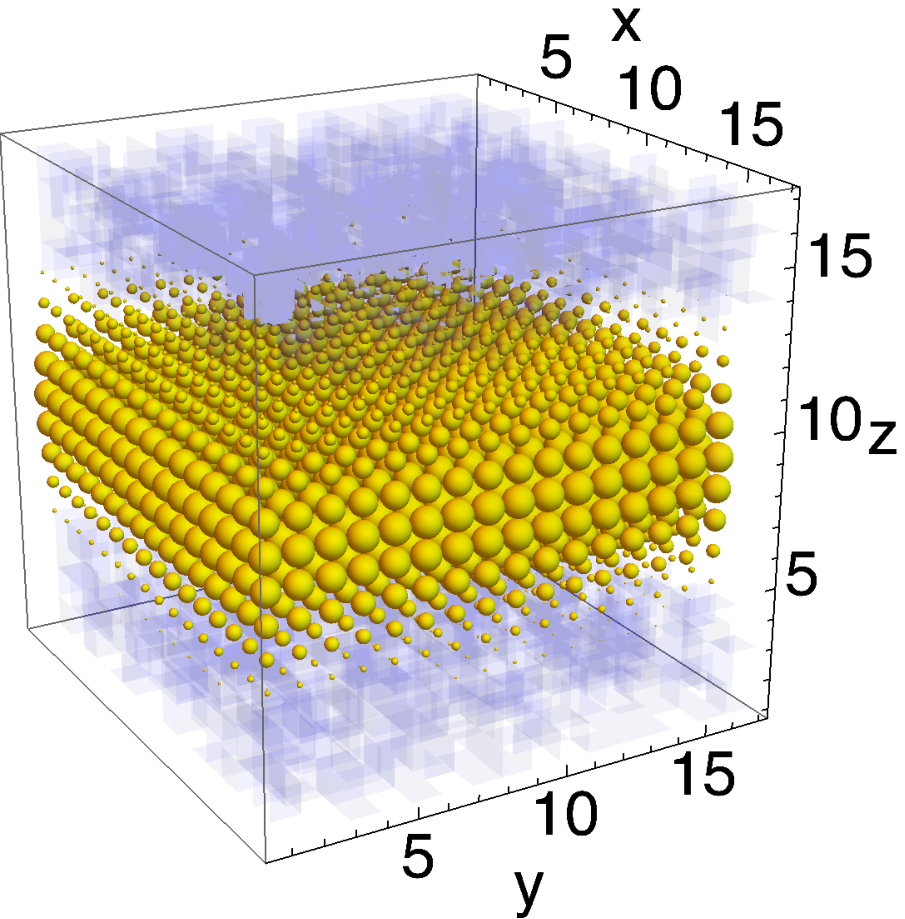}
    }
    \caption{
    Snapshots of the spinon probability density (yellow spheres) and the corresponding vison background (in blue) from our Monte Carlo simulations. The spinon wavefunction and depletion region are generally spherical (top left panel). At low temperatures, due to finite-sized effects, we observe an instability first to a cylindrical shape (top right panel) and finally to a planar slab (bottom panel) before the visons are completely expelled from the system. The temperatures for these visualizations are $T/\eta_s = 1.0 \times 10^{-4}$, $5.2 \times 10^{-5}$, and $4.0 \times 10^{-5}$ for the sphere, cylinder, and slab, respectively.}
    \label{fig:instabilities}
\end{figure}

At the lowest temperatures, the vison depletion region discussed in Sec.~\ref{sec:spinonvison} becomes comparable in size with the simulation cell. The system displays finite-sized instabilities which alter the shape of the depletion region from a sphere to a cylinder and eventually to a planar slab (see Fig.~\ref{fig:instabilities}) as the temperature is lowered. 

This is observed in the numerics and can be understood by effective free energy considerations like those of Eq.~\eqref{eqn:free_en_spinon}. Here, we only provide an analytical calculation of the transition temperatures at saddle-point approximation and in the limit $\xi \gg \xi_0$ and $\Delta_v=0$, for convenience. 
In the case of a spherical region, we minimize Eq.~\eqref{eqn:free_en_spinon} with respect to $\xi$ to obtain
\begin{eqnarray}
    \xi_*^s &=& \left(\frac{\pi \eta_s}{2T \ln 2 }\right)^{1/5},
\\
F(\xi_*^s) &=& \frac{5}{3}\pi^2 \left(\frac{2T \ln 2}{\pi \eta_s}\right)^{2/5}
\, . 
    \label{eqn:xisphere}
\end{eqnarray}
In the case of a cylindrical region, Eq.~\eqref{eqn:free_en_spinon} needs to be modified to 
\begin{equation}
    \label{eqn:free_en_cyl}
    F_{c}(\xi) = 
    \frac{j_0^2 \eta_s}{(\xi+\xi_0)^2} 
    + 
    \pi {\xi}^{2}L T 
    \mathrm{ln}\,[1+\exp(-\beta {{{\Delta }}}_{\text{v}})]
    \, , 
\end{equation}
shown for completeness including $\xi_0$ and $\Delta_v$.
Minimizing again with respect to $\xi$ gives 
\begin{eqnarray}
    \xi_*^c &=& \left(\frac{j_0^2 \eta_s}{\pi T L \ln 2 }\right)^{1/4},
\\
F_c(\xi_*^c) &=& 2 j_0^2 \eta_s \left(\frac{\pi LT \ln 2}{j_0^2 \eta_s}\right)^{1/2} 
\, . 
\label{eqn:xicyl}
\end{eqnarray}

Finally, in the case of a planar slab depletion region, %
\begin{equation}
    \label{eqn:free_en_sl}
    F_{sl}(\xi) = 
    \frac{\pi^2 \eta_s}{4(\xi+\xi_0)^2} 
    + 
    2 \xi L^2 T 
    \mathrm{ln}\,[1+\exp(-\beta {{{\Delta }}}_{\text{v}})]
    \, , 
\end{equation}
shown again for completeness including $\xi_0$ and $\Delta_v$. We minimize with respect to $\xi$ to obtain 
\begin{eqnarray}
    \xi_*^{sl} &=& \left(\frac{\pi^2 \eta_s}{4 T L^2 \ln 2 }\right)^{1/3}
    \nonumber, \\ 
    F_{sl}(\xi_*^{sl}) &=& \frac{3}{4}\pi^2 \eta_s \left( \frac{4L^2 T \ln 2}{\pi^2 \eta_s} \right)^{2/3}
    \label{eqn:xisl}
    \, . 
\end{eqnarray}

The stable equilibrium geometry of the vison depletion region is determined by the lowest free energy, and transitions between them occur at the crossing temperature where Eq.~\eqref{eqn:xisphere} equals Eq.~\eqref{eqn:xicyl}: 
\begin{equation}
    \label{eqn:tsc}
    T_{\mathrm{sc}} = \left(\frac{5 \pi}{6j_0}\right)^{10} \frac{16 \pi}{\ln 2} \frac{\eta_s}{L^5}
    \simeq 1.62 \times 10^{-4} \eta_s
    \, ,
\end{equation}
and where Eq.~\eqref{eqn:xicyl} equals Eq.~\eqref{eqn:xisl}: 
\begin{equation}
    \label{eqn:tcsl}
    T_{\mathrm{csl}} = \left(\frac{2 j_0}{3 \pi^{1/3}}\right)^{6} \frac{16 \pi}{\ln 2} \frac{\eta_s}{L^5}
    \simeq 1.19 \times 10^{-4} \eta_s
    \, ,
\end{equation}
where indeed $T_{\mathrm{sc}} > T_{\mathrm{csl}}$. Both instabilities take place within the physical range of parameters, namely, well before the vison density falls below the $1/{\rm Volume}$ limit. 

\begin{figure}[ht!]
    \centering
    \includegraphics[width=\linewidth]{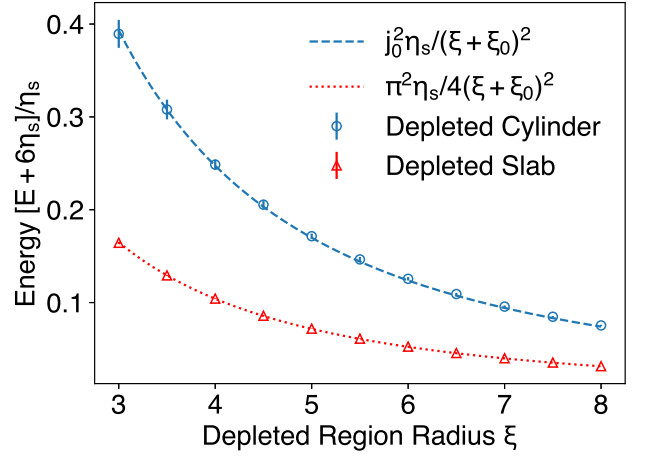}
    \caption{\label{fig:cylinderxi0}
    Ground state energy of a spinon in an artificial vison background generated with depleted cylindrical and planar slab regions spanning the entire system, as a function of the linear size $\xi$ of the vison depleted region. The data are averaged over $500$ realizations of systems of size $L^3=20^3$. The error bars are shown (and generally smaller than the symbol size). The dashed lines are the best fits to the energy of an infinite three-dimensional (3D) potential well of corresponding shape but with linear size $\xi + \xi_0$ (the radius of the cylinder and the thickness of the slab); this gives $\xi_0=1.835(3)$ and $=1.869(2)$ for the cylinder and the slab, respectively. The vertical axis has been shifted and rescaled for convenience.
    }
\end{figure}

\begin{figure}[ht!]
    \centering
    \includegraphics[width=\linewidth]{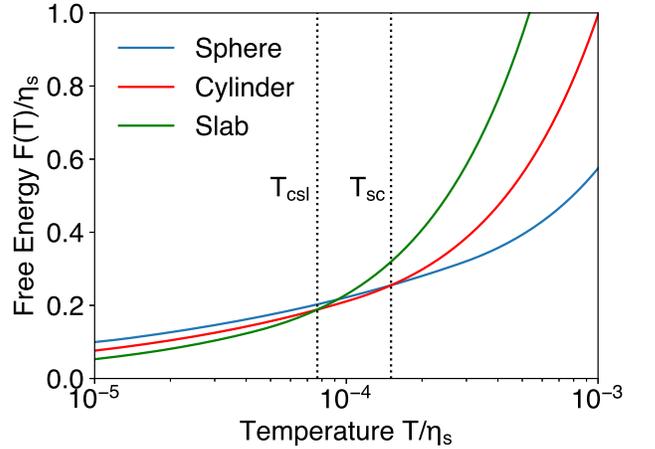}
    \caption{\label{fig:finsizeinst}
    Effective free energy plots for temperatures between $T/\eta_s = 10^{-5}$ and $10^{-3}$ for the three geometries of the vison depletion region: spherical, cylindrical, and planar slab. Here, we use the improved free energies that account for the penetration depth $\xi_0$ obtained from the fits in Figs.~\ref{fig:MainDiagram} and \ref{fig:cylinderxi0} for system size $L^3=16^3$. 
    The free energy diagram predicts a transition from sphere to cylinder at $T_{\mathrm{sc}}/\eta_s = 1.50 \times 10^{-4}$ and a transition from cylinder to strip at $T_{\mathrm{csl}}/\eta_s = 7.67 \times 10^{-5}$.}
\end{figure}

The same calculations can be carried out including a finite $\xi_0$ fitting parameter as in the main text. The equations are less transparent, and for convenience, we limit ourselves to illustrate the results graphically. 
Following the same procedure as in Fig.~\ref{fig:MainDiagram}, we fit the $\xi_0$ values for the cylindrical and planar slab geometries in Fig.~\ref{fig:cylinderxi0}.
With the respective fit values $\xi_0=1.814(0), \, 1.835(3), \, 1.869(2)$ (sphere/cylinder/slab), the improved saddle-point free energies are plotted in Fig.~\ref{fig:finsizeinst} as a function of temperature, showing the finite-sized instabilities at the crossing points $T/\eta_s = 1.50 \times 10^{-4}$ (sphere to cylinder) and $T/\eta_s = 7.67 \times 10^{-5}$ (cylinder to planar slab). 

The saddle-point approximation discussed in this appendix paints a picture that is consistent with the one empirically observed in our MC simulations, which show a transition from a sphere to a cylinder around $T_{\mathrm{sc}}/\eta_s \simeq 6.1 \times 10^{-5}$ (see also Fig.~\ref{fig:MainDiagram}) and a transition from cylinder to planar slab around $T_{\mathrm{csl}}/\eta_s \simeq 4.5 \times 10^{-5}$.
%
%
\newpage
\bibliographystyle{unsrt_abbrv_custom}
\bibliography{References.bib}{}
\normalsize

\end{document}